# Abundant Atmospheric Methane from Volcanism on Terrestrial Planets Is Unlikely and Strengthens the Case for Methane as a Biosignature

Nicholas Wogan[1,2] , Joshua Krissansen-Totton[2,3], and David C. Catling[1,2]
[1] Dept. Earth and Space Sciences, University of Washington, Seattle, WA, USA; wogan@uw.edu
[2] Virtual Planetary Laboratory, University of Washington, Seattle, WA, USA
[3] Dept. of Astronomy and Astrophysics, University of California, Santa Cruz, CA, USA

## Abstract

The disequilibrium combination of abundant methane and carbon dioxide has been proposed as a promising exoplanet biosignature that is readily detectable with upcoming telescopes such as the James Webb Space Telescope. However, few studies have explored the possibility of nonbiological $CH_4$ and $CO_2$ and related contextual clues. Here we investigate whether magmatic volcanic outgassing on terrestrial planets can produce atmospheric $CH_4$ and $CO_2$ with a thermodynamic model. Our model suggests that volcanoes are unlikely to produce $CH_4$ fluxes comparable to biological fluxes. Improbable cases where volcanoes produce biological amounts of $CH_4$ also produce ample carbon monoxide. We show, using a photochemical model, that high abiotic $CH_4$ abundances produced by volcanoes would be accompanied by high CO abundances, which could be a detectable false-positive diagnostic. Overall, when considering known mechanisms for generating abiotic $CH_4$ on terrestrial planets, we conclude that observations of atmospheric $CH_4$ with $CO_2$ are difficult to explain without the presence of biology when the $CH_4$ abundance implies a surface flux comparable to modern Earth's biological $CH_4$ flux. A small or negligible CO abundance strengthens the $CH_4+CO_2$ biosignature because life readily consumes atmospheric CO, while reducing volcanic gases likely cause CO to build up in a planet's atmosphere. Furthermore, the difficulty of volcanically generated $CH_4$-rich atmospheres suitable for an origin of life may favor alternatives such as impact-induced reducing atmospheres.

*Unified Astronomy Thesaurus concepts:* Biosignatures (2018); Astrobiology (74); Planetary atmospheres (1244); Volcanoes (1780)

## 1. Introduction

Large telescopes will soon be used to search for biogenic waste gases in exoplanet atmospheres. Oxygen is the most extensively studied biosignature gas (Meadows 2017; Meadows et al. 2018). Although many studies have proposed ways of identifying scenarios where nonliving processes might mimic life by producing oxygen (i.e., false positives; Domagal-Goldman et al. 2014; Tian et al. 2014; Wordsworth & Pierrehumbert 2014; Harman et al. 2015; Luger & Barnes 2015; Schwieterman et al. 2019), the circumstances are unusual and contextual clues can distinguish abiotic scenarios (Meadows et al. 2018).

However, even when life is present, oxygen biosignatures may be uncommon. Oxygenic photosynthesis is a complex metabolism that only evolved once on Earth (Fischer et al. 2016). Additionally, oxygen was slow to accumulate in the Earth's atmosphere (Lyons et al. 2014), and other planets may have low $O_2$ concentrations for billions of years despite having oxygenic photosynthetic life if there are large oxygen sinks (Claire et al. 2006). Accumulation of oxygen may be especially challenging on planets orbiting M-dwarf stars due to their low visible photon flux, which potentially limits primary production (Lehmer et al. 2018).

One alternative to detecting oxygen-rich planets like the modern Earth is to look for methane on planets like the Archean Earth. Before the rise of oxygen, methanogenic life could have sustained a methane-rich atmosphere, which could be detected with remote spectroscopy (Schindler & Kasting 2000; Kasting & Catling 2003).

Recently, Krissansen-Totton et al. (2018b) proposed a criterion for methane biosignatures: finding abundant $CH_4$ in the presence of $CO_2$ (abbreviated $CH_4+CO_2$). This combination is compelling if the $CH_4$ mixing ratio is greater than 0.1% because it is difficult to explain such an abundance with the short atmospheric lifetime of $CH_4$ in terrestrial atmospheres and nonbiological methane sources such as serpentinization (Krissansen-Totton et al. 2018b). This 0.1% threshold value is for planets that orbit stars like the Sun and must be adjusted for different stellar types. For example, planets orbiting M-stars typically receive less near-UV radiation than planets orbiting Sun-like stars, resulting in different photochemistry that promotes the buildup of $CH_4$ (Segura et al. 2005; Grenfell et al. 2007, 2014; Rugheimer et al. 2015; Rugheimer & Kaltenegger 2018). Krissansen-Totton et al. (2018b) argued that the $CH_4$ biosignature is strengthened by a low CO abundance because volcanoes that produce $CH_4$ should also likely generate CO. Additionally, living planets might have low CO because microbes consume CO (Kharecha et al. 2005); coupled ecosystem-planetary models of the early Earth suggest atmospheric $CO/CH_4$ ratios declined dramatically with the emergence of chemoautotrophic ecosystems (Sauterey et al. 2020).

Exploring false positives for methane biosignatures is timely. Biogenic $O_2$ or $O_3$ detections with upcoming telescopes, such as the James Webb Space Telescope (JWST), will be extremely difficult (Barstow & Irwin 2016; Krissansen-Totton et al. 2018a; Fauchez et al. 2019; Lustig-Yaeger et al. 2019; Wunderlich et al. 2020), whereas $CH_4+CO_2$ biosignatures are more readily detectable. Indeed, an Archean-Earth-like CH$_4$+CO$_2$ biosignature is potentially detectable on the planet TRAPPIST-1e, with just 10 transits (Krissansen-Totton et al. 2018a). Thus, exploration of potential methane biosignature false positives and their contextual discriminants is needed.

The literature exploring false positives for methane biosignatures has primarily focused on CH$_4$ generation in deep-sea serpentinizing hydrothermal vents. Guzmán-Marmolejo et al. (2013) estimated a maximum CH$_4$ surface flux of 0.18 Tmol yr$^{-1}$ (6.8 × 10$^8$ molecules cm$^{-2}$ s$^{-1}$) from hydrothermal vents for planets with the same mass as Earth. Additionally, Krissansen-Totton et al. (2018b) used Monte Carlo simulations to estimate a probability distribution for maximum abiotic CH$_4$ production from this process. They suggest that >10 Tmol CH4 yr$^{-1}$ is highly unlikely. These estimated maximum fluxes are small compared to modern Earth's biological CH$_4$ flux of 30 Tmol yr$^{-1}$.

However, investigations of abiotic CH$_4$ on Earth suggest that these estimates of abiotic CH$_4$ from hydrothermal vents are potentially unrealistically large. Serpentinization reactions involving water and ultramafic oceanic crust generate H$_2$; then, purportedly, H$_2$ might react with inorganic carbon in hydrothermal systems to generate CH$_4$. Krissansen-Totton et al. (2018b) and Guzmán-Marmolejo et al. (2013) both estimated abiotic CH$_4$ fluxes, assuming efficient reactions between H$_2$ and inorganic carbon. However, laboratory experiments have shown that, uncatalyzed, this reaction is extremely slow at hydrothermal vent temperatures and pressures preventing chemical equilibrium on timescales of at least months (Reeves & Fiebig 2020). Additionally, various lines of evidence suggest that much of the CH$_4$ observed in deep-sea hydrothermal vent waters is ultimately from biology (Reeves & Fiebig 2020). Furthermore, lifeless planets without silica-secreting organisms should have high ocean-water SiO$_2$ concentrations, which suppresses the H$_2$ and therefore abiotic CH$_4$ produced from serpentinization (Tutolo et al. 2020).

Impacts can likely generate abiotic CH$_4$ (Zahnle et al. 2020), although impact-generated CH$_4$ is only probable early in a solar system's lifetime. The cratering record on the Moon shows that Earth's impact flux decreased dramatically by 3.5 Ga (Marchi et al. 2014). Thus, extra-solar systems that are several billion years old are probably unlikely to have abiotic CH$_4$ from this source.

Here we investigate another potential false-positive for the CH$_4$+CO$_2$ biosignature: magma-sourced volcanic outgassing (i.e., not metamorphic). Negligible CH$_4$ has been observed in gases emitted by magmatic volcanoes on Earth (Catling & Kasting 2017; Reeves & Fiebig 2020), although it has not been investigated whether substantial CH$_4$ is feasible for volcanoes in vastly different thermodynamic regimes. We simulate outgassing speciation for a range of magma temperatures, outgassing pressures, oxygen fugacities, volatile composition, and variable partitioning between subaerial and submarine volcanism. We examine whether volcanoes can produce CH$_4$ fluxes comparable to biological fluxes. Using a photochemical model, we also investigate the atmospheric composition of hypothetical planets by reducing volcanic gases to see whether volcanic CH$_4$ coincides with large atmospheric CO, which could be a detectable false-positive marker.

## 2. Methods

### 2.1. Model for Calculating Volcanic Outgassing Speciation

Below, we describe our model for predicting the gases produced by an erupting mantle-sourced volcano. We follow Gaillard & Scaillet (2014) and solve for the gas–gas and gas–melt equilibrium in a C–O–H system. Our model differs from Gaillard & Scaillet (2014) because we do not consider nitrogen or sulfur species. Despite these differences, we obtain similar results to calculations made in Gaillard & Scaillet (2014). We have also validated our code against the work of Liggins et al. (2020) and Ortenzi et al. (2020), which have independently constructed similar outgassing models. Our Python code is published as open-source software on the GitHub page https://github.com/Nicholaswogan/VolcGases.

Figure 1 shows a highly schematic conceptualization of volcanic degassing typical of low-viscosity magma. Gas bubbles form in the magma when molecules like H$_2$O and CO$_2$ are exsolved. Within the gas bubbles, reactions drive the system to chemical equilibrium. The oxygen fugacity ($f_{O_2}$) of the gas bubble is controlled by equilibrium with the oxygen fugacity of the magma (e.g., Kadoya et al. 2020). Gases bubbles are released from the magma and enter the overlying atmosphere or ocean.

A mathematical model describes the volatiles in gas bubbles and magma. The amount of carbon and hydrogen that are exsolved by the magma into bubbles is governed by the solubility of CO$_2$ and H$_2$O, which we calculate with the solubility relations for mafic magmas described in Iacono-Marziano et al. (2012):

$$\ln(x_{CO_2}) = x_{H_2O} d_{H_2O} + a_{CO_2} \ln(P_{CO_2}) + S_1, \quad (1)$$

$$\ln(x_{H_2O}) = a_{H_2O} \ln(P_{H_2O}) + S_1. \quad (2)$$

Here $x_{CO_2}$ and $x_{H_2O}$ are mol fractions of CO$_2$ and H$_2$O in the magma, respectfully. Additionally, $P_{CO_2}$ and $P_{H_2O}$ are the partial pressure of CO$_2$ and H$_2$O in gas bubbles suspended in the magma. The other terms in Equations (1) and (2) are solubility parameters with values shown in Table 1, except $S_1$ and $S_2$, which are further described in Appendix A.1. We use solubility relations appropriate for mafic magmas because rocky planets and moons in our solar system usually have basaltic crusts, suggesting that mafic magma is common to most terrestrial bodies.

Volatile mol fractions (e.g., $x_{H_2O}$) can be converted to mass fractions with the formula

$$m_i = \frac{x_i \mu_i}{\mu_{magma}} \quad (3)$$

Here $m_i$ is the mass fraction, $\mu_i$ is the volatile's molar mass, and $i$ can be either H$_2$O or CO$_2$. Table 1 gives the units of each term.

We assume that after the hot gas exsolves from the magma into bubbles, it achieves thermodynamic equilibrium from the reactions

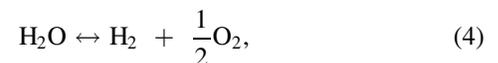

$$H_2O \leftrightarrow H_2 + \frac{1}{2}O_2, \quad (4)$$

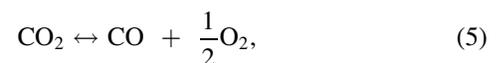

$$CO_2 \leftrightarrow CO + \frac{1}{2}O_2, \quad (5)$$

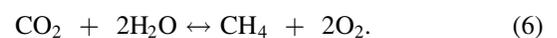

$$CO_2 + 2H_2O \leftrightarrow CH_4 + 2O_2. \quad (6)$$

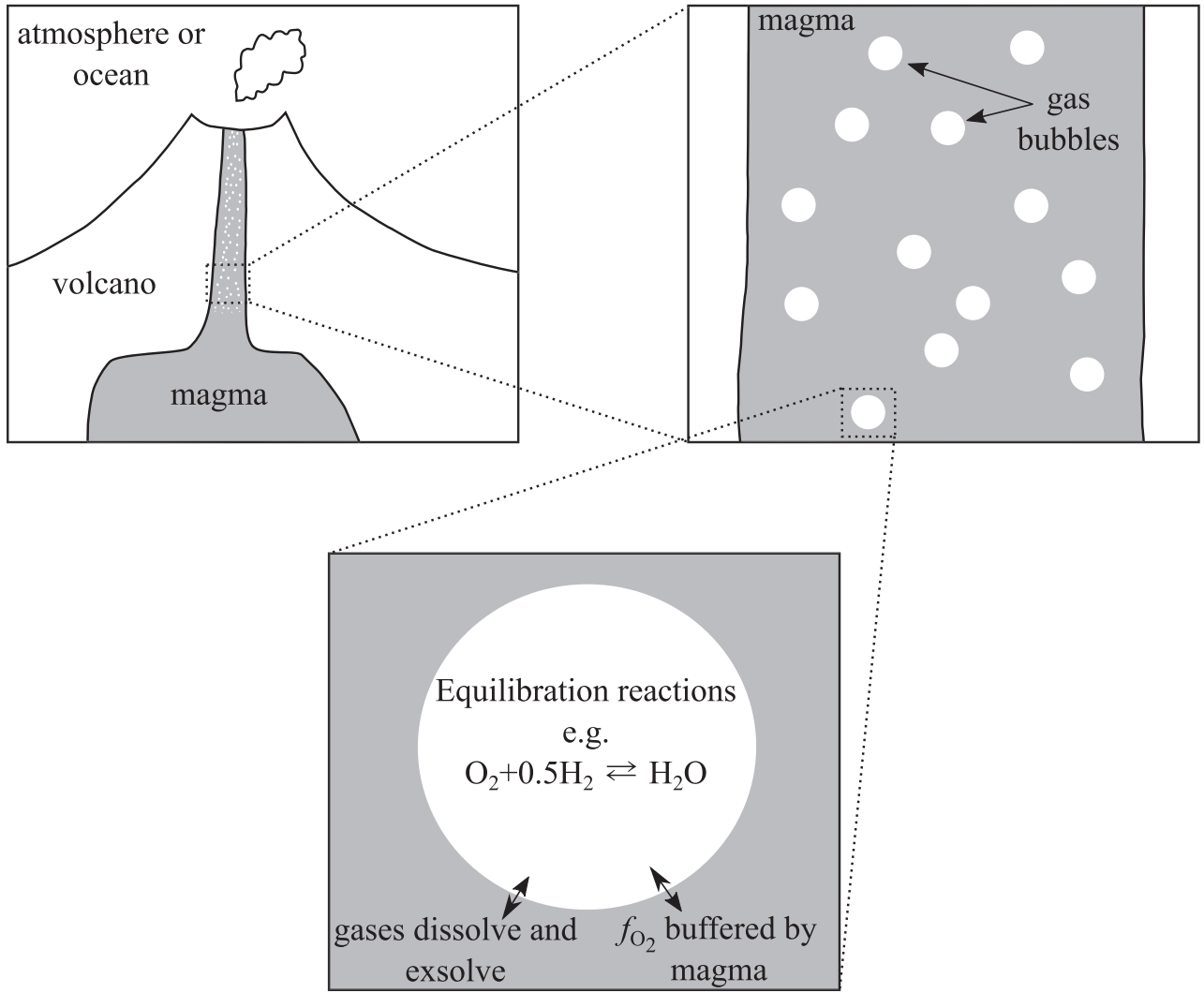

**Figure 1.** Qualitative sketch of degassing typical of low-viscosity magma (e.g., Hawaiian volcanoes). Here a gas bubble reaches thermal and chemical equilibrium with a melt (no crystals are present). Note, degassing can occur in many different ways depending on magma viscosity and volatile content (Gonnermann & Manga 2013).

At thermodynamic equilibrium, the ratios of the fugacities of volatile species (denoted $f_i$) are related to the equilibrium constant corresponding to each chemical reaction. We assume that we can replace fugacities with partial pressures (denoted $P_i$). This approximation is reasonable for the temperatures and pressures involved in volcanic outgassing (Holland 1984). Thus,

$$K_1 = \frac{f_{H_2} f_{O_2}^{0.5}}{f_{H_2O}} \approx \frac{P_{H_2} f_{O_2}^{0.5}}{P_{H_2O}}, \quad (7)$$

$$K_2 = \frac{f_{CO} f_{O_2}^{0.5}}{f_{CO_2}} \approx \frac{P_{CO} f_{O_2}^{0.5}}{P_{CO_2}}, \quad (8)$$

$$K_3 = \frac{f_{CH_4} f_{O_2}^2}{f_{CO_2} f_{H_2O}^2} \approx \frac{P_{CH_4} f_{O_2}^2}{P_{CO_2} P_{H_2O}^2}. \quad (9)$$

We calculate equilibrium constants (e.g., $K_1$) using the NASA thermodynamic database (Burcat & Ruscic 2005). We assume that the gas is thermally and chemically coupled to the magma so that the oxygen fugacity ($f_{O_2}$) of the gas is set by the oxygen fugacity of magma, as observed (Symonds et al. 1994). So far, we have seven unknowns ($x_{CO_2}$, $x_{H_2O}$, $P_{CO_2}$, $P_{H_2O}$, $P_{CO}$, $P_{H_2}$, $P_{CH_4}$) and only five equations. To close the system, we add three more equations and one more unknown. The first equation requires that the partial pressures sum to the total pressure:

$$P_{H_2} + P_{H_2O} + P_{CO} + P_{CO_2} + P_{CH_4} = P. \quad (10)$$

The final two equations are atom conservation equations for carbon and hydrogen:

$$\frac{m_{CO_2}^{tot} \mu_{magma}}{\mu_{CO_2}} = \frac{P_{CO_2} + P_{CO} + P_{CH_4}}{P} \alpha_{gas} + (1 - \alpha_{gas}) x_{CO_2}, \quad (11)$$

$$\frac{m_{H_2O}^{tot} \mu_{magma}}{\mu_{H_2O}} = \frac{P_{H_2O} + P_{H_2} + 2P_{CH_4}}{P} \alpha_{gas} + (1 - \alpha_{gas}) x_{H_2O}. \quad (12)$$

Equations (11) and (12) state that the total moles of either carbon or hydrogen should be equal to the moles of either

Table 1
Model Constants and Variables

|  | Constant or variable | Value | Units | Definition |
|---|---|---|---|---|
| Constants | $d_{H_2O}$ | 2.3 | ... | Solubility constant[a] |
|  | $a_{CO_2}$ | 1 | ... | Solubility constant[a] |
|  | $a_{H_2O}$ | 0.54 | ... | Solubility constant[a] |
|  | $S_1$ | ... | ... | Solubility constant[a] |
|  | $S_2$ | ... | ... | Solubility constant[a] |
|  | $\mu_{magma}$ | 64.52 | $\frac{g\ magma}{mol\ magma}$ | Molar mass of magma[b] |
|  | $\mu_{H_2O}$ | 18.02 | $\frac{g\ H_2O}{mol\ H_2O}$ | Molar mass of $H_2O$ |
|  | $\mu_{CO_2}$ | 44.01 | $\frac{g\ CO_2}{mol\ CO_2}$ | Molar mass of $CO_2$ |
|  | $K_1$ | $e^{-29755/T+6.55}$ | $bar^{0.5}$ | Equilibrium constant[c] |
|  | $K_2$ | $e^{-33979/T+10.42}$ | $bar^{0.5}$ | Equilibrium constant[c] |
|  | $K_3$ | $e^{-96444/T+0.22}$ | ... | Equilibrium constant[c] |
| Input | $P$ | ... | bar | Total pressure of degassing |
|  | $T$ | ... | K | Temperature of magma and gas |
|  | $f_{O_2}$ | ... | bar | Oxygen fugacity of the magma |
|  | $m_{CO_2}^{tot}$ | ... | $\frac{g\ CO_2}{g\ gas\ and\ magma}$ | Mass fraction $CO_2$ in magma before degassing |
|  | $m_{H_2O}^{tot}$ | ... | $\frac{g\ H_2O}{g\ gas\ and\ magma}$ | Mass fraction $H_2O$ in magma before degassing |
| Output | $x_{H_2O}$ | ... | $\frac{mol\ H_2O}{mol\ magma}$ | Mol fraction of $H_2O$ in the magma after degassing |
|  | $x_{CO_2}$ | ... | $\frac{mol\ CO_2}{mol\ magma}$ | Mol fraction of $CO_2$ in the magma after degassing |
|  | $P_{H_2O}$ | ... | bar | Partial pressure of $H_2O$ |
|  | $P_{CO_2}$ | ... | bar | Partial pressure of $CO_2$ |
|  | $P_{H_2}$ | ... | bar | Partial pressure of $H_2$ |
|  | $P_{CO}$ | ... | bar | Partial pressure of CO |
|  | $P_{CH_4}$ | ... | bar | Partial pressure of $CH_4$ |
|  | $\alpha_{gas}$ | ... | $\frac{mol\ gas}{mol\ gas\ and\ magma}$ | Mol fraction in gas phase |

**Notes.**
[a] From Iacono-Marziano et al. (2012). See Appendix A.1 to calculate $S_1$ and $S_2$.
[b] Molar mass of Mount Etna magma.
[c] Calculated from the NASA thermodynamic database (Burcat & Ruscic 2005).

element in the gas phase plus the moles in the magma. Here $\alpha_{gas}$ is the final unknown. It is the total moles in the gas phase divided by the total moles in the gas and magma combined. See Appendix A.2 for a full derivation of Equations (11) and (12).

Given a gas and magma temperature ($T$), pressure ($P$), oxygen fugacity ($f_{O_2}$), and the total mass fraction (or mol fraction) of $CO_2$ and $H_2O$ in the magma ($m_{CO_2}^{tot}$, and $m_{H_2O}^{tot}$), Equations (1), (2), (7)–(12) are a system of eight equations and eight unknowns ($x_{CO_2}$, $x_{H_2O}$, $P_{CO_2}$, $P_{H_2O}$, $P_{CO}$, $P_{H_2}$, $P_{CH_4}$, $\alpha_{gas}$). We solve this system of equations numerically with the Scipy Python package.

The solution to this system of equilibrium equations provides an estimate of the amount of each volatile species in gas bubbles in magma immediately before the gas leaves the magma. We assume bubbles remain in thermodynamic equilibrium with the surrounding melt until they are released into the overlying atmosphere or ocean, and volatile speciation does not continue to evolve upon release. This does not exactly reflect real degassing. Observed outgassing chemistry suggests that volcanic gas re-equilibrates to temperatures slightly lower than the magma as the gas leaves the magma and is no longer chemically buffered by it (Oppenheimer et al. 2018; Moussallam et al. 2019; Kadoya et al. 2020). We do not capture this complexity in the main text, although in Appendix A.4 we investigate the closed system re-equilibration of volcanic gases and show that this process does not change our conclusions.

Once the unknowns are solved for, they can be used to calculate the gas production (i.e., the moles of gas produced per kilogram of magma erupted):

$$q_i = 10^3 \left( \frac{\alpha_{gas}}{\mu_{magma}(1-\alpha_{gas})} \right) \frac{P_i}{P}. \quad (13)$$

Here $q_i$ is the gas production of species $i$ in mol gas kg$^{-1}$ magma. Calculating $q_i$ is useful because it is related to the flux $F_i$ of gas $i$ to the atmosphere by the magma production rate:

$$F_i = q_i Q_m. \quad (14)$$

Here $Q_m$ is the magma production rate in kg magma yr$^{-1}$ and $F_i$ is in mol yr$^{-1}$.

Several authors have shown that degassing can be affected by graphite saturation of magma (Hirschmann & Withers 2008) or by the solubility of CO, $CH_4$, and $H_2$ in magma (Hirschmann et al. 2012; Ardia et al. 2013; Wetzel et al. 2013). The gas speciation model described previously does not account for these processes. However, in Appendix A.3, we introduce a more complex model that accounts for graphite saturation and CO, $CH_4$, and $H_2$ solubility, and show that this model produces very similar results to the simplified model described here.

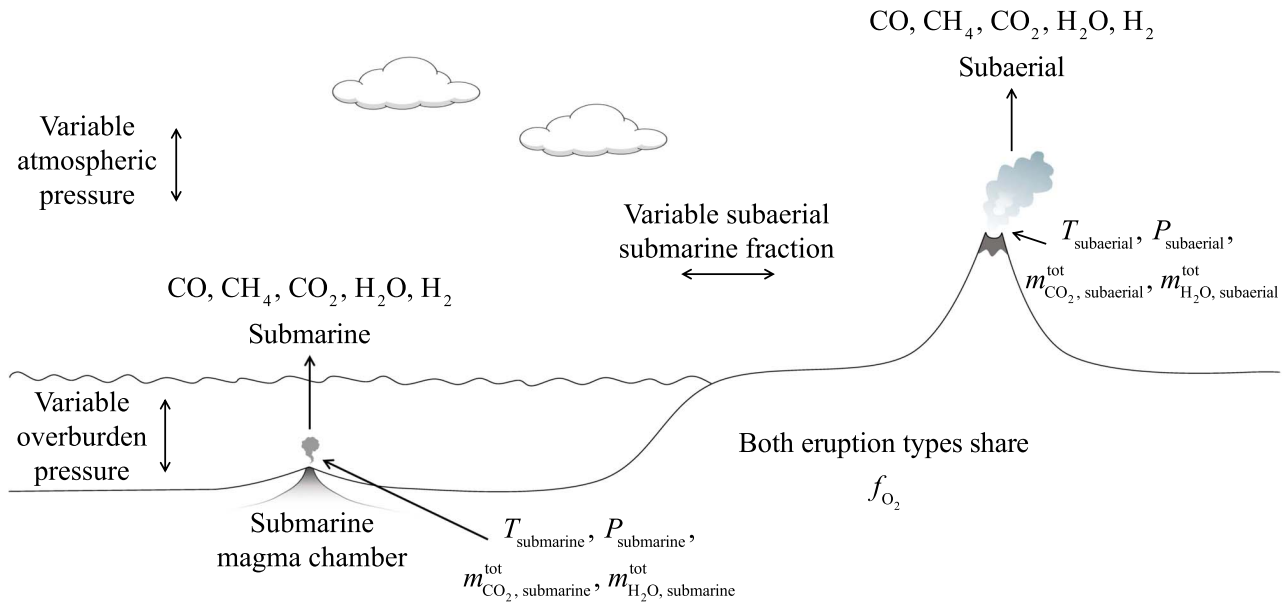

**Figure 2.** Illustration of the parameters considered in the Monte Carlo simulations.

**Table 2**
Monte Carlo Sampling Distributions

| Variable | Low | High | Sampling method | Justification |
|---|---|---|---|---|
| $T_{\text{submarine}}$ | 873 K | 1973 K | Linear uniform | Range of submarine magma temperatures observed on Earth[a] |
| $T_{\text{subaerial}}$ | 873 K | 1973 K | Linear uniform | Range of subaerial magma temperatures observed on Earth[a] |
| $P_{\text{submarine}}$ | 100 bar | 1000 bar | Linear uniform | Degassing pressure at 1 km to 10 km ocean depth[b] |
| $P_{\text{subaerial}}$ | 0.001 bar | 100 bar | $\log_{10}$ uniform | Rough range of subaerial degassing pressure in solar system |
| $m^{\text{tot}}_{CO_2,\text{ submarine}}$ | $10^{-5}$ | $10^{-2}$ | $\log_{10}$ uniform | Approx. $CO_2$ mass fraction range in Earth magma (Wallace 2005; Wallace et al. 2015; Anderson & Poland 2017; le Voyer et al. 2019) |
| $m^{\text{tot}}_{CO_2,\text{subaerial}}$ | $10^{-5}$ | $10^{-2}$ | $\log_{10}$ uniform | Approx. $CO_2$ mass fraction range in Earth magma (Wallace 2005; Wallace et al. 2015; Anderson & Poland 2017; le Voyer et al. 2019) |
| $m^{\text{tot}}_{H_2O,\text{ submarine}}$ | $10^{-5}$ | $10^{-1}$ | $\log_{10}$ uniform | $H_2O$ mass fraction range for Earth submarine outgassing (Wallace et al. 2015) |
| $m^{\text{tot}}_{H_2O,\text{ subaerial}}$ | $10^{-5}$ | $10^{-1}$ | $\log_{10}$ uniform | $H_2O$ mass fraction range for Earth subaerial outgassing (Wallace et al. 2015) |
| $f_{O_2}$ | FMQ-4 | FMQ+5 | $\log_{10}$ uniform | Oxygen fugacity of most reducing Martian meteorite (Catling & Kasting 2017) to most oxidized magma on Earth (Stamper et al. 2014)[c] |
| $X$ | 0 | 1 | Linear uniform | 0% to 100% subaerial volcanism |

**Notes.**
[a] Coldest rhyolite magma and hottest komatiites magmas (Huppert et al. 1984).
[b] Assumes Earth's gravity. The solubility of $H_2O$ in magma does not allow for significant $CH_4$ degassing at pressures greater than 1000 bar, equivalent to a depth of 10 km.
[c] FMQ is the fayalite-magnetite-quartz mineral redox buffer. See Chapter 7 in Catling & Kasting (2017) for a description of mineral redox buffers. We use the parameterization for the FMQ buffer defined by Wones & Gilbert (1969). This parameterization has only been experimentally validated to 1400 K (O'Neill 1987), but we extrapolate using the parameterization to 1973 K.

## 2.2. Monte Carlo Simulations

We investigate volcanic false positives to the $CH_4+CO_2$ biosignature on two types of worlds: an Earth-like world with subaerial and submarine outgassing (Figure 2) and an ocean world with only submarine outgassing. For each type of planet, we search for false-positive scenarios by calculating volcanic outgassing speciation with a wide range of input parameters.

To explore volcanism on Earth-like planets, we calculate outgassing speciation 10,000 times. For each calculation, we sample either uniform or $\log_{10}$-uniform distributions (see Table 2) of 10 parameters: $T_{\text{submarine}}$, $P_{\text{submarine}}$, $m^{\text{tot}}_{CO_2,\text{ submarine}}$, $m^{\text{tot}}_{H_2O,\text{ submarine}}$, $T_{\text{subaerial}}$, $P_{\text{subaerial}}$, $m^{\text{tot}}_{CO_2,\text{ subaerial}}$, $m^{\text{tot}}_{H_2O,\text{ subaerial}}$, $f_{O_2}$, and $X$. The width of each uniform sampling distribution is given and explained in Table 2. We use inputs with subscripts "subaerial" to calculate subaerial volcanic speciation and inputs with subscripts "submarine" to calculate submarine volcanic speciation, and then we combine the results of each calculation with the formula

$$n_i = \frac{P_{i,\text{ subaerial}}}{P_{\text{subaerial}}} X + \frac{P_{i,\text{ submarine}}}{P_{\text{submarine}}} (1-X). \quad (15)$$

Here $n_i$ is the mixing ratio of averaged outgassed volatiles of species $i$ produced by the combination of subaerial and submarine volcanoes and $X$ is the fraction of subaerial volcanism ($0 < X < 1$). Also, $P_{i,\text{subaerial}}$ and $P_{i,\text{submarine}}$ are the partial pressure of species $i$ in subaerial and submarine outgassing, respectively.

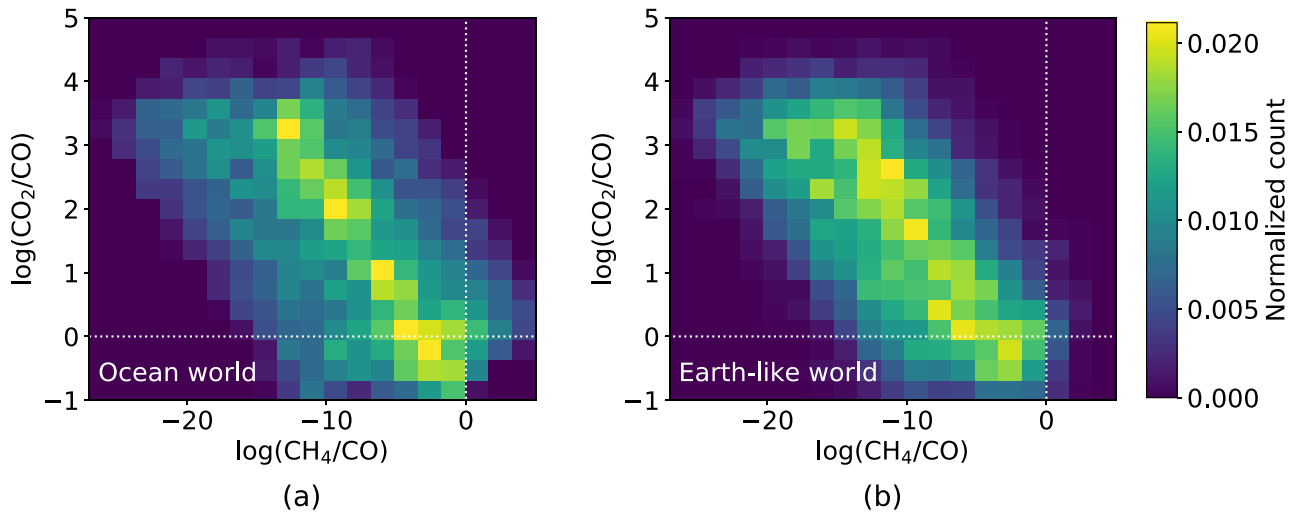

**Figure 3.** Results of the Monte Carlo simulation described in Section 2.2. (a) and (b) show normalized count as a function of $\log(CH_4/CO)$ and $\log(CO_2/CO)$ for an ocean world and Earth-like world, respectively. The white dotted lines indicate where $CH_4/CO = 1$ and $CO_2/CO = 1$. For almost all calculated gas speciations, $CO_2$ and CO are much more abundant than $CH_4$.

To investigate volcanism on an ocean world, we also calculate outgassing speciation 10,000 times. For each calculation, we sample either uniform or $\log_{10}$-uniform distributions of inputs $T_{\text{submarine}}$, $P_{\text{submarine}}$, $m^{\text{tot}}_{CO_2,\text{ submarine}}$, $m^{\text{tot}}_{H_2O,\text{ submarine}}$, and $f_{O_2}$, with ranges defined and justified in Table 2.

### 2.3. Photochemical Modeling: Uninhabited Anoxic Ocean World with Reducing Volcanic Gases

We further investigate the $CH_4+CO_2$ biosignature by modeling the atmospheric composition of hypothetical uninhabited ocean worlds with reducing volcanic gases. We consider planets orbiting the Sun and a late M star—the latter because planets orbiting M-dwarfs are the most feasible targets for near-term telescopes like JWST (Barstow & Irwin 2016). Additionally, we simulate ocean worlds because ocean-bottom degassing is most thermodynamically prone to produce $CH_4$, as revealed by our Monte Carlo simulations and previous studies (French 1966; Kasting & Brown 1998; see Section 4.1.1 for further discussion).

To simulate atmospheres on uninhabited planets, we use the 1-D photochemical model contained within the open-source software package *Atmos*. *Atmos* is derived from a model originally developed by the Kasting group (Pavlov et al. 2001), and versions of this code have been used to simulate the Archean and Proterozoic Earth atmosphere (Zahnle et al. 2006), Mars (Zahnle et al. 2008; Smith et al. 2014; Sholes et al. 2019), and exoplanet atmospheres (Harman et al. 2015; Schwieterman et al. 2019).

## 3. Results

### 3.1. Monte Carlo Simulations

Figure 3 shows joint distributions of gas ratios $CH_4/CO$ and $CO_2/CO$ from the Monte Carlo simulation described in Section 2.2. These results suggest that for most combinations of parameters, volcanoes are most likely to produce more $CO_2$ than CO, and negligible $CH_4$, which is the case for the modern Earth (Catling & Kasting 2017). About 7% and 2% of calculations produce more $CH_4$ than CO for ocean worlds and Earth-like worlds, respectfully. In the vast majority of cases, either CO or $CO_2$ is the dominant carbon-bearing species.

Figures 4(a) and (b) show $CH_4$ production from the Monte Carlo simulations in terms of mol $CH_4$ kg$^{-1}$ magma. To give a sense of the gas fluxes implied by these $CH_4$ productions, we multiply the distributions in Figures 4(a) and (b) by the magma production rate of modern Earth of $9 \times 10^{13}$ kg yr$^{-1}$ (Crisp 1984), which gives the gas fluxes shown in Figures 4(c) and (d), respectively. About 0.1% of calculations predict more than 10 Tmol $CH_4$ yr$^{-1}$ for both Earth-like worlds and ocean worlds. This small fraction suggests that for modern Earth magma production rates, volcanoes are unlikely to produce $CH_4$ fluxes comparable to modern Earth's biological flux of 30 Tmol yr$^{-1}$ (Hauglustaine et al. 2007).

Magma production rates larger than modern Earth's increase the probability that volcanic fluxes of $CH_4$ become comparable to biological $CH_4$ fluxes. For example, the early Archean Earth could have had magma production rates up to about 25 times modern Earth's (Sleep & Zahnle 2001). Such a magma production rate would shift the distributions in Figures 4(c) and (d) to larger values by a factor of 25 (or in $\log_{10}$-space, by a factor of 1.4). In this case, ~2% of calculations (for either Earth-like world or ocean world) would predict more than 10 Tmol $CH_4$ yr$^{-1}$.

Crucially, large $CH_4$ fluxes should almost always coincide with even larger CO fluxes (horizontal axis in Figure 3). Therefore, the unlikely cases where volcanoes mimic biological $CH_4$ fluxes can be identified by detecting abundant CO in a planet's atmosphere. We further investigate CO as a $CH_4+CO_2$ biosignature discriminant using a photochemical model in the following section.

### 3.2. Photochemical Modeling: Uninhabited Anoxic Ocean World with Reducing Volcanic Gases

We use the *Atmos* photochemical model to simulate the potential observable gas abundances of uninhabited Earth-sized ocean worlds with reducing volcanic gases. We consider such planets because they are the most prone to mimic biology by producing volcanic $CH_4$ (see Section 4.1.1 for more details). Our hypothetical planets have 1 bar $N_2$ dominated atmospheres, 400 bars of ocean water, magma degassing at 1473 K, and mantle redox states of FMQ-4. Here FMQ is the fayalite-magnetite-quartz buffer, which is a synthetic reference $f_{O_2}$ value at fixed temperature-pressure conditions. Additionally, we assume that the magma contains 0.1 wt% CO$_2$ and 1 wt% H$_2$O. Our assumed H$_2$O concentration is comparable to those observed in submarine hot-spot magmas (0.2 to 1.5 wt%; Wallace et al. 2015); however, the CO$_2$ concentration we assume is slightly lower (Anderson & Poland 2017). Given these inputs, our speciation model (Section 2.1) predicts gas production from erupted magma of $q_{H_2} = 4.36 \times 10^{-2}$ mol gas/kg magma, $q_{CO} = 1.29 \times 10^{-2}$ mol gas/kg magma, and $q_{CH_4} = 7.39 \times 10^{-3}$ mol gas/kg magma.

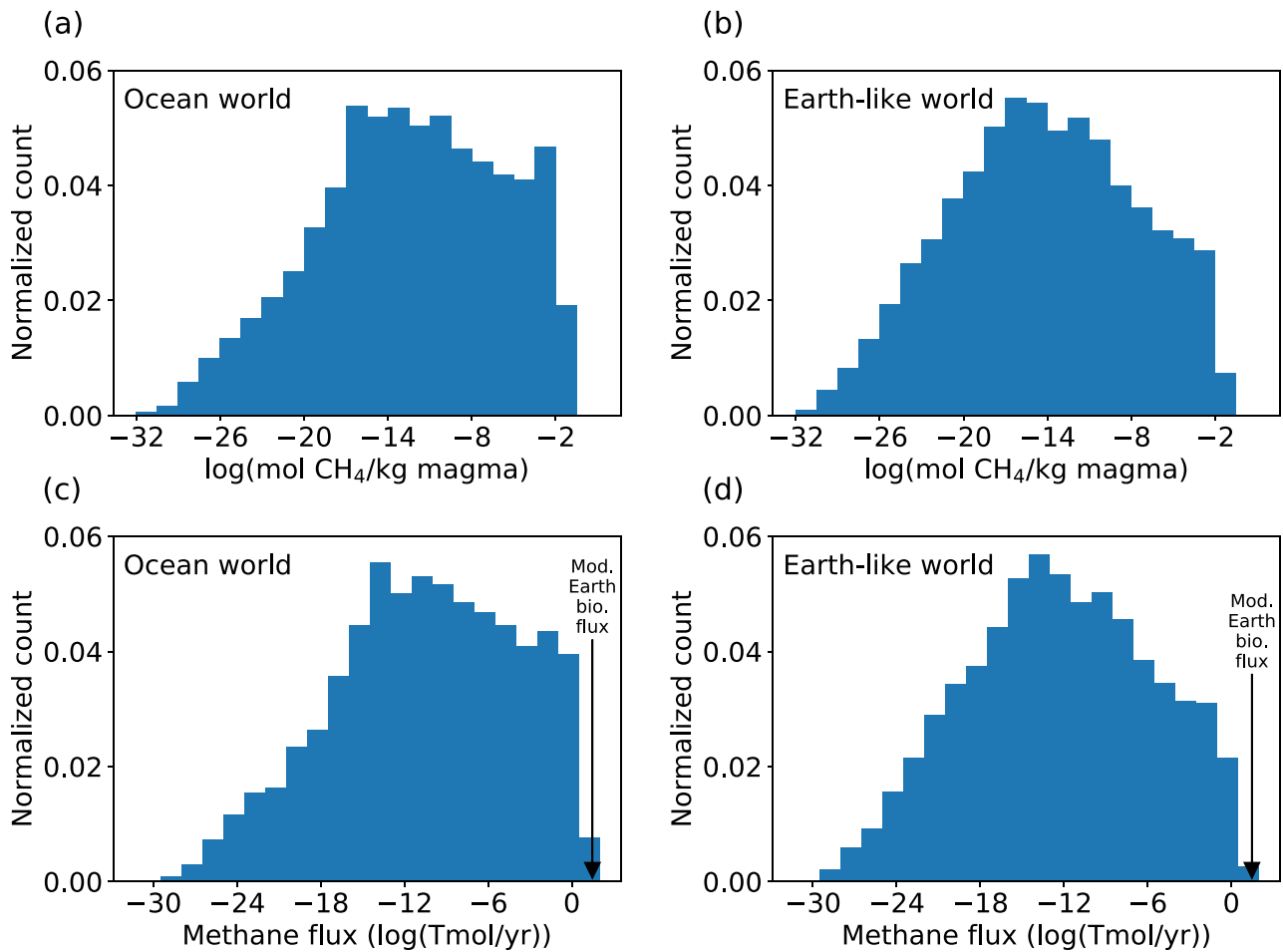

**Figure 4.** Normalized count of methane production (mol gas/kg magma) for (a) ocean worlds and (b) Earth-like worlds. Distributions were calculated by sampling the ranges in Table 2. Multiplying Earth's magma production rate of $9 \times 10^{13}$ kg magma yr$^{-1}$ by (a) and (b) gives the methane fluxes in (c) and (d), respectively. For modern Earth's magma production rate, volcanoes are likely to produce negligible CH$_4$.

The magnitude of gas fluxes to the atmosphere resulting from chemically reducing volcanism depends on the magma production rate (Equation (14)). We consider magma production rates between about $10^{-3}$ and $10^2$ Earth's modern magma production rate of $9 \times 10^{13}$ kg magma yr$^{-1}$ (Crisp 1984).

For each magma production rate, we calculate the outgassing flux of CH$_4$, H$_2$, and CO and set these fluxes as lower boundary conditions to the *Atmos* photochemical model. (The outgassing model also gives CO$_2$ and H$_2$O fluxes, but we do not use them in our photochemical modeling.) *Atmos* only allows fixed CO$_2$ mixing ratios and not CO$_2$ fluxes, so we consider cases with low and high CO$_2$ (100 ppm and 10%). Additionally, we set the deposition velocity of CO to $10^{-8}$ cm s$^{-2}$ to reflect the abiotic uptake of CO by the ocean (Kharecha et al. 2005). All other boundary conditions are specified in Appendix B. Given volcanic outgassing fluxes and other boundary conditions, *Atmos* calculates the mixing ratios of all species when the atmosphere is at photochemical equilibrium.

Figure 5 shows the photochemical modeling results of reducing volcanic gases on an uninhabited Earth-sized ocean world orbiting the Sun. Figure 5(a) assumes that the atmosphere has 100 ppmv CO$_2$, while Figure 5(b) assumes that atmospheric CO$_2$ is 10%. Carbon monoxide and methane are more abundant in the model with more CO$_2$ because CO$_2$ shields the lower atmosphere from hydoxyl (OH) production from water photolysis. In anoxic atmospheres, OH is a significant sink for both CO and CH$_4$ through the reactions CO$_2$ + OH → CO$_2$ + H and CH$_4$ + OH → CH$_3$ + H$_2$O. OH is generated primarily from H$_2$O photolysis (H$_2$O + $h\nu$ [$\lambda$ < 200 nm] → OH + H), but CO$_2$ shields H$_2$O from photolysis in model runs with 10% CO$_2$, thus limiting the CH$_4$ and CO destruction from OH. Also, CH$_4$ is more abundant in atmospheres with more CO$_2$ because CO$_2$ shields CH$_4$ from direct photolysis in cases when CO$_2$ is >200 times as abundant as CH$_4$. This factor of ~200 comes from comparing Ly$\alpha$ ($\lambda$ = 121.6 nm) CO$_2$ and CH$_4$ cross sections. Ly$\alpha$ is the portion of the UV spectrum primarily responsible for photolyzing CH$_4$.

Figure 5 suggests that reducing volcanic gases on an ocean world orbiting a Sun-like star will only mimic biological CH$_4$

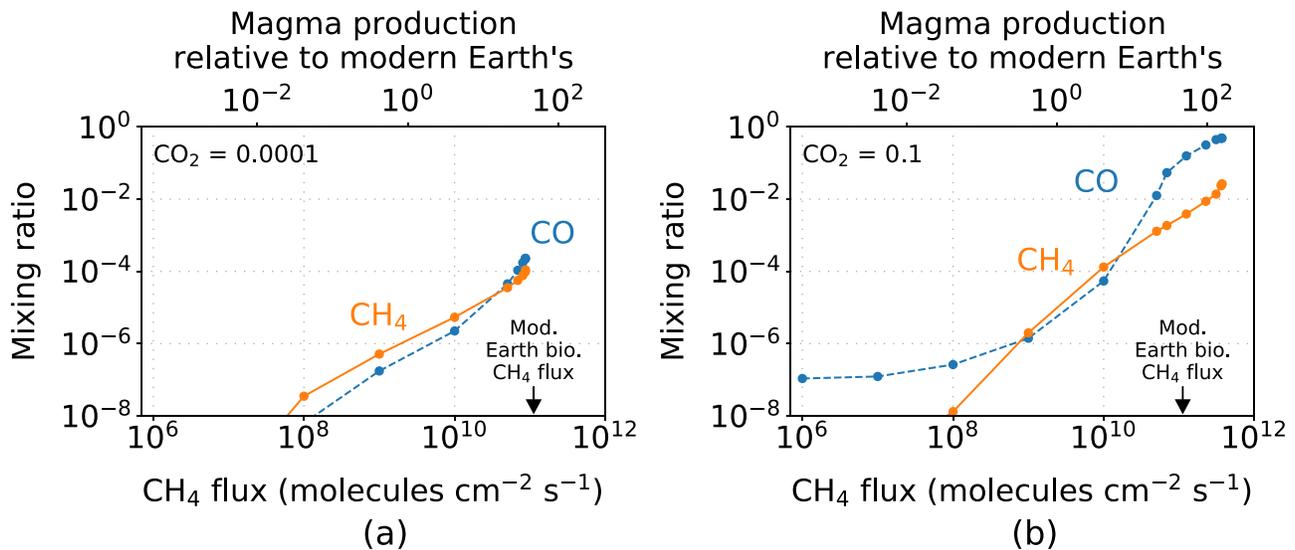

**Figure 5.** Atmospheric mixing ratios of CO and $CH_4$ as a function of magma production rate relative to modern Earth's (or $CH_4$ flux) on an anoxic ocean world with reducing volcanic gases orbiting a Sun-like star. (a) and (b) are identical model runs, except (a) assumes a constant atmospheric $CO_2$ mixing ratio of 0.0001, and (b) assumes a constant atmospheric $CO_2$ mixing ratio of 0.1. Modern Earth's biological $CH_4$ flux is indicated on the horizontal axes. Archean Earth-like $CH_4$ fluxes and abundances are only mimicked by volcanoes for magma production rates >10 times modern Earth's. Such false-positive cases can be distinguished from biology because the CO abundance exceeds the $CH_4$ abundance, which would likely not be the case for an inhabited planet.

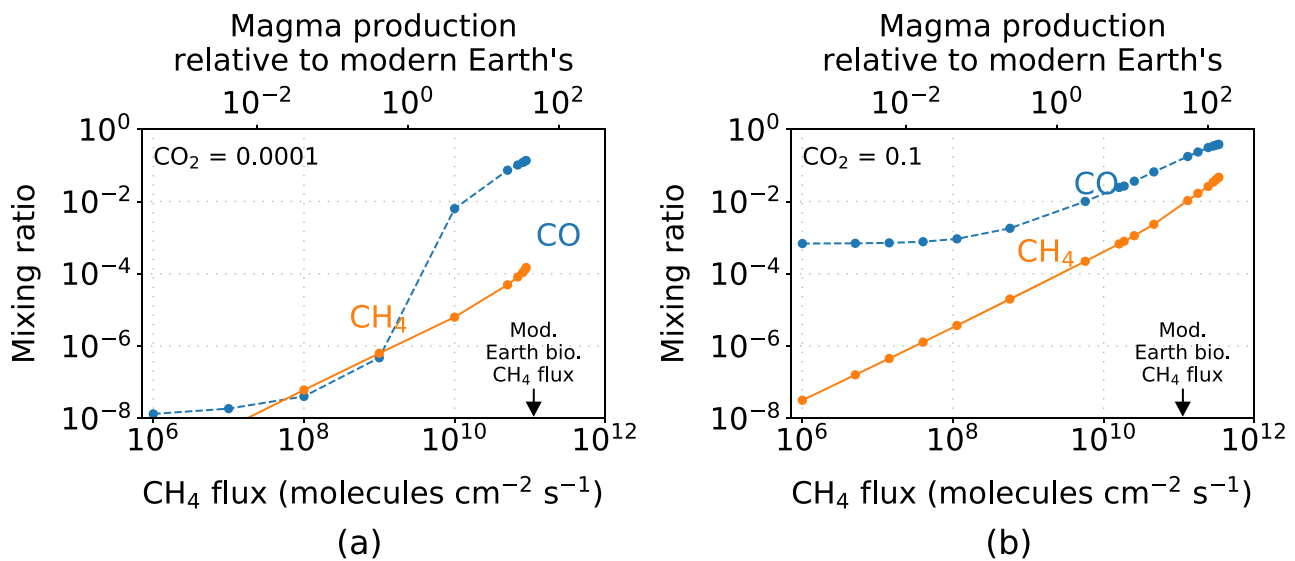

**Figure 6.** Identical to Figure 5, except for a planet that orbits an M8V star instead of a Sun-like star.

fluxes and abundances for large magma production rates. Volcanism can generate Earth's modern biological $CH_4$ flux when the magma production rate is ∼50 times modern Earth's (Figure 5). In this case, the photochemical model predicts an atmospheric $CH_4$ abundance between 0.01% and 0.3%, depending on the $CO_2$ mixing ratio. Such $CH_4$ abundances are similar to the 0.01% to 1% expected in the early Archean Earth atmosphere (Catling & Zahnle 2020). In contrast, magma production rates comparable to the modern Earth's result in a $CH_4$ flux of $2.4 \times 10^9$ molecules $cm^{-2}$ $s^{-1}$ (0.64 Tmol $yr^{-1}$) and $CH_4$ abundances <30 ppm, which are likely to be considered abiotic levels in an anoxic atmosphere.

Figure 6 shows the CO and $CH_4$ mixing ratios on an Earth-sized ocean world with reducing volcanic gases orbiting a cold M star. CO and $CH_4$ are more abundant on the ocean world orbiting the M star compared to the ocean world orbiting a Sun-like star (Figure 5). This is because M8V stars have a low flux of near-ultraviolet radiation compared to Sun-like stars. The low near-ultraviolet flux reduces OH produce from $H_2O$ photolysis, thus allowing for relatively high CO and $CH_4$ concentrations.

One consequence of M-dwarf photochemistry is a higher likelihood of Archean Earth-like $CH_4$ abundances on uninhabited planets with reducing gases from volcanism. Figure 6 shows that modern Earth magma production rates can result in $CH_4$ abundances up to 0.01%, which is comparable to what is expected in the Archean atmosphere.

Potential $CH_4$ biosignature false positives from reducing volcanic gases might be discriminated from inhabited worlds using observations of CO. For planets orbiting Sun-like stars (Figure 5) or M stars (Figure 6), the CO abundance is higher than the $CH_4$ abundance in every case that is a potential outgassing false-positive. Some authors have argued that a large CO abundance is unlikely on an inhabited planet, because atmospheric CO should be readily consumed by biology (Krissansen-Totton et al. 2018a). Conversely, Schwieterman et al. (2019) has demonstrated hypothetical cases where large CO can coincide with biology in an anoxic atmosphere. We further discuss CO as a false-positive discriminant in Section 4.2.

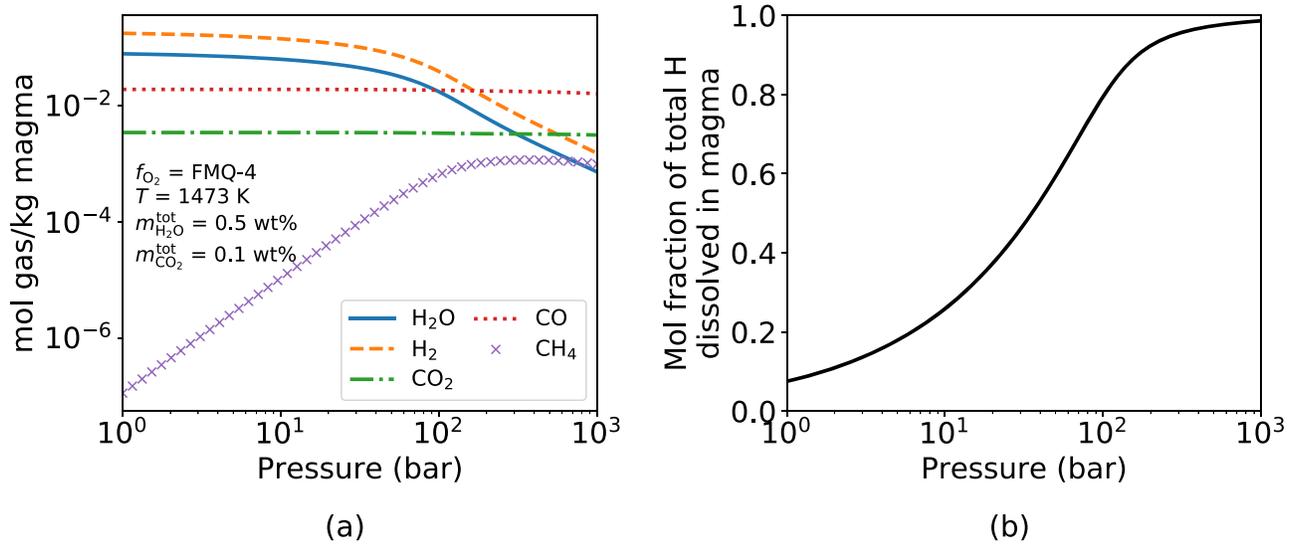

**Figure 7.** (a) Modeled gas speciation as a function of pressure. (b) Mole fraction of total hydrogen dissolved in the magma as a function of pressure. Model assumes $f_{O_2}$ = FMQ-4, $T = 1473$ K, $m_{H_2O}^{tot} = 0.5$ wt%, and $m_{CO_2}^{tot} = 0.1$ wt%. Methane becomes more prevalent in volcanic gases at higher pressures but asymptotes because hydrogen dissolves into the magma, reducing the total amount of H-bearing volatiles released from the magma.

## 4. Discussion

### 4.1. The Reasons Why Volcanoes Produce Little CH₄

Our modeling results show that for modern Earth magma production rates, volcanic fluxes of reducing gases are unlikely to produce more than 1 Tmol CH₄ yr⁻¹, even in an extreme case (Figure 4). This flux is relatively small compared to the flux of other volcanic gases on modern Earth. For example, Earth's modern volcanoes produce about 7.5 Tmol $CO_2$ yr⁻¹ and 95 Tmol $H_2O$ yr⁻¹ (Catling & Kasting 2017, p. 203). There are three main reasons why the outgassing model predicts little CH₄, which we explore further in the following discussion.

#### 4.1.1. Volcanoes Produce Little CH₄ because of Water Solubility in Magma

One reason for small CH₄ outgassing is the high solubility of water in magma at high pressures. Consider Equation (9), which can be re-arranged as follows:

$$\frac{P_{CH_4}}{P_{CO_2}} = \frac{K_3 P_{H_2O}^2}{f_{O_2}^2}. \quad (16)$$

The ratio $P_{CH_4}/P_{CO_2}$ in a gas bubble in magma is directly proportional to $P_{H_2O}^2$ within that bubble. Generally speaking, $P_{H_2O}$ increases as the total pressure of degassing increases because all partial pressures must sum to the total pressure (Equation (10)). For example, subaerial degassing at ~1 bar will have a relatively small $P_{H_2O}$ and thus a small $P_{CH_4}/P_{CO_2}$ ratio. On the other hand, submarine degassing at ~400 bar should have a larger $H_2O$ partial pressure and thus a larger $P_{CH_4}/P_{CO_2}$ ratio. Here the equilibrium constant and oxygen fugacity have extremely weak pressure dependencies (i.e., they are effectively constant as degassing pressure changes).

Figure 7(a) shows modeled gas speciation for highly reducing volcanism ($f_{O_2}$ = FMQ-4) as a function of pressure. For small pressures (<100 bar), CH₄ increases with increasing pressure and then asymptotes for pressures >100 bar.

CH₄ asymptotes because of the high solubility of water in magma at high pressure. High pressures dissolve a large fraction of the total available hydrogen as $H_2O$ into the magma, which is shown in Figure 7(b). Dissolving a large amount of $H_2O$ into the magma limits the amount of hydrogen available in the gas phase for making H-bearing species, like CH₄, $H_2O$, and $H_2$.

In summary, high pressure is in some ways thermodynamically favorable for making methane because $P_{CH_4}/P_{CO_2} \propto P_{H_2O}^2$, but it is also unfavorable because high pressure dissolves a large fraction of the available hydrogen in the magma as $H_2O$. Limited amounts of hydrogen in gas bubbles result in small amounts of CH₄ produced.

Kasting & Brown (1998) used Equation (16) to argue that ~1% of the carbon outgassed by submarine volcanoes should be CH₄ for magma with $f_{O_2}$ = FMQ. They assumed that $P_{H_2O} \approx P$, the total pressure. This assumption is valid for oxidized subaerial volcanoes because ~90% of the gas exsolved by Earth's subaerial volcanoes is $H_2O$ (Catling & Kasting 2017, p. 203). However, $P_{H_2O} < P$ for submarine volcanoes because of the high-water solubility in magma at high pressure. Our outgassing model, which accounts for water's solubility in magma, produces negligible methane.

Li & Lee (2004) also predict abundant CH₄ produced by subaerial and submarine volcanoes (their Figure 5). However, they calculated equilibrium constants in units of bars and then used units of Pascals for equilibrium chemistry calculations. The result was that they calculated speciation for pressures a factor 10,000 times greater than reported. For example, we were able to reproduce their subaerial outgassing case (their Figure 5(a)) by assuming $P = 10,000$ bar and not the $P = 1$ bar total pressure they intended. Additionally, like Kasting & Brown (1998), they did not account for the high solubility of $H_2O$ in magma at high pressure. Their methods assume the total hydrogen outgassed for submarine volcanoes is the same as the total hydrogen outgassed by subaerial volcanoes. This should not be the case, because at high pressure water dissolves in magma and is unavailable for making H-bearing gas species (Figure 7(b)).

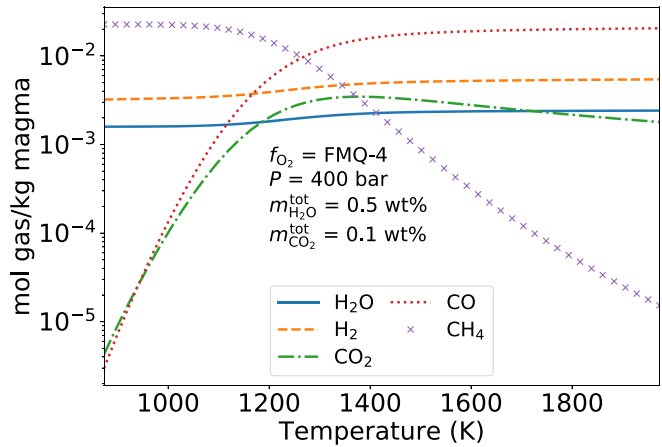

**Figure 8.** Modeled volcanic outgassing speciation as a function of temperature. Model assumes $f_{O_2}$ = FMQ-4, $P$ = 400 bar, $m_{H_2O}^{tot}$ = 0.5 wt%, and $m_{CO_2}^{tot}$ = 0.1 wt%. $CH_4$ is more thermodynamically favorable at lower degassing temperatures.

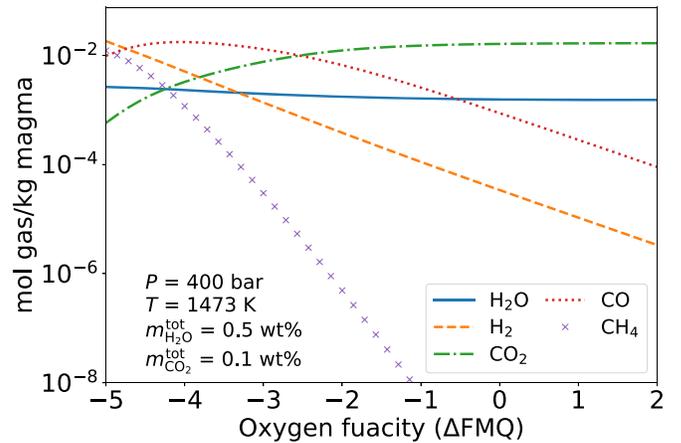

**Figure 9.** Modeled volcanic outgassing speciation as a function of oxygen fugacity. Model assumes $P$ = 400 bar, $T$ = 1473 K, $m_{H_2O}^{tot}$ = 0.5 wt%, and $m_{CO_2}^{tot}$ = 0.1 wt%. Methane is most favorable at low oxygen fugacity.

The pressure dependence of volcanic outgassing has implications for planetary atmospheres generally (Gaillard & Scaillet 2014). Thin atmospheres will allow substantial degassing of both carbon and hydrogen bearing species. However, planets with thick atmospheres or large global oceans will have volcanic degassing dominated by $CO_2$ and CO, and almost no hydrogen bearing species. The overburden pressure where C-bearing species dominate depends primarily on the un-degassed concentrations of $H_2O$ and $CO_2$ in the magma. In Figure 7, $CO_2$ and CO overwhelm H-bearing species at ~1000 bar for initial volatile concentrations of $m_{CO_2}^{tot}$ = 0.1% and $m_{H_2O}^{tot}$ = 0.5%. In contrast, Figure 8 in Gaillard & Scaillet (2014) illustrates a case with less volatiles ($m_{CO_2}^{tot}$ = 0.007% and $m_{H_2O}^{tot}$ = 0.03%) where C-bearing species eclipse H-bearing species at ~1 bar.

*4.1.2. Volcanoes Produce Little $CH_4$ because Magma Is Hot*

Relatively little $CH_4$ is produced by volcanoes because $CH_4$ is generally not thermodynamically favorable at typical magma degassing temperatures. Figure 8 shows gas speciation as a function of temperature for a submarine outgassing case. For these chosen inputs, $CH_4$ is the dominant carbon-bearing species for $T < 1200$ K. Mid-ocean ridge basalts (MORB) are about 2/3 of total magma produced on Earth (Crisp 1984). MORB magma erupts at temperatures between 1473 and 1650 K (Scheidegger 1973) and are thus in a temperature regime where $CH_4$ is unfavorable, even from more reducing volcanism.

On the other hand, magma from arc volcanoes is generally much colder than MORB magma. Moussallam et al. (2019) report magma temperatures for many arc volcanoes (their Table S3), the coldest of which are 1123 K. Thus, it does seem possible for magma to be cold enough for $CH_4$ to be the dominant carbon-bearing outgassed species from an extremely reducing volcano with $f_{O_2}$ = FMQ-4.

Recall that large magma production rates (~30x modern) are required for volcanoes to produce $CH_4$ fluxes compared to biological ones (Figure 5). It seems unlikely that planets with large magma production rates will have magma temperatures cold enough to produce plentiful $CH_4$. For example, the Archean Earth may have had a larger magma production rate than the modern Earth because the Earth's mantle was hotter in the distant past (Sleep & Zahnle 2001). The hotter Archean mantle resulted in the eruption of ~1800 K komatiite magmas (Huppert et al. 1984) or possibly only ~1600 K (McKenzie 2020). Such hot magma degassing is unfavorable for methane (Figure 8).

*4.1.3. Volcanoes Produce Little $CH_4$ because Very Low Oxygen Fugacity Is Required*

The final reason why volcanic $CH_4$ is unlikely on terrestrial planets is because very low $f_{O_2}$ is required to make abundant methane. Figure 9 shows gas speciation as a function of oxygen fugacity for submarine volcanism. For these assumed inputs, methane is a substantial fraction of outgassed species for $f_{O_2} <$ FMQ-3, and at FMQ-5 (roughly equivalent to the quartz-fayalite-iron buffer), half the carbon is converted to $CH_4$, while the other half is CO. Most degassing on Earth occurs at approximately $f_{O_2}$ = FMQ (Catling & Kasting 2017, p. 208), but magma spans FMQ-4 to FMQ+5 (Stamper et al. 2014). Additionally, the oxygen fugacity of Martian meteorites ranges between FMQ and FMQ-3.7 (Catling & Kasting 2017, p. 363). Therefore, the $f_{O_2} <$ FMQ-3 required for plentiful $CH_4$ outgassing is at the extremes of the oxygen fugacities observed for Earth and Mars.

Astronomical observations and geochemical experiments suggest Earth-sized planets should generally have relatively oxidized magmas. Doyle et al. (2019) spectroscopically measured the oxygen fugacity of material polluting the surface of several white dwarfs. Their observations suggest that rocky exoplanets are likely to have similar oxygen fugacities to Earth and Mars. Additionally, high pressure experiments suggest that the upper mantles of Earth-sized planets should self-oxidize by iron oxide disproportionation to roughly FMQ during the magma-ocean phase, early in a planet's life (Armstrong et al. 2019).

*4.2. Carbon Monoxide as a Methane Biosignature Discriminant*

CO-consuming life evolved very early on Earth (Adam et al. 2018) and is a relatively simple metabolism. Therefore, it seems possible that life on other planets will evolve to consume CO. Planets with atmospheric $CH_4+CO_2$ produced by life might also have relatively small amounts of atmospheric CO because of CO consumers. Consequentially, the presence of abundant CO along with $CH_4$ can discriminate abiotic situations.

Monte Carlo simulations show that volcanoes should almost always produce more CO than $CH_4$ (Figure 3). Additionally, photochemical modeling (Figures 5 and 6) suggests that CO should build up in the atmospheres of uninhabited planets with reducing submarine volcanic gases. Thus atmospheric $CO_2+CH_4$ produced by volcanoes is likely accompanied by a large CO concentration. This is distinct from an inhabited world, which can have lower CO concentrations due to CO-consuming life.

However, the mere presence of large atmospheric CO is not a definitive sign of an uninhabited planet with reducing volcanic gases (Schwieterman et al. 2019). This is because there are limits to how quickly gases can be transported from the atmosphere into the ocean where they can be consumed by life (Kharecha et al. 2005). For example, consider a planet with a very large volcanic CO flux (e.g., 100x modern). CO could build up in this planet's atmosphere even if CO consumers were present in an ocean because CO transport from the atmosphere to the ocean would not be sufficient to maintain low atmospheric CO.

In summary, the $CH_4+CO_2$ biosignature is most compelling when the CO abundance is low or negligible because a lack of CO potentially implies the presence of CO-consuming biology. In comparison, atmospheric $CH_4+CO_2$ and large CO is ambiguous, and can either be explained by reducing volcanic gases or by an inhabited world that is unable to sequester atmospheric CO.

JWST might be able to put a tentative upper limit on atmospheric CO. Krissansen-Totton et al. (2018a) simulated JWST retrievals of TRAPPIST-1e with an atmospheric composition similar to the Archean Earth containing 10 ppbv CO. Their synthetic retrieval suggested CO was below 652 ppmv with 90% confidence after 10 transits. CO constraints could be improved by co-adding more transits and positive CO detections may also be possible with JWST (Wunderlich et al. 2020).

However, even if observational CO constraints are poor, it may still be possible to say something about the abiotic or biotic origin of atmospheric $CH_4$. Reducing gases from volcanism is unlikely to mimic the modern biological $CH_4$ flux of 30 Tmol yr$^{-1}$ (Section 4.1). Additionally, serpentinization is unlikely to produce 30 Tmol $CH_4$ yr$^{-1}$, and impact-generated $CH_4$ might be distinguished with system age (Krissansen-Totton et al. 2018b). Therefore, JWST observations of atmospheric $CH_4+CO_2$ would be challenging to explain without the presence of biology regardless of atmospheric CO, as long as the $CH_4$ abundance implies a surface flux similar to the modern Earth's.

### 4.3. $CH_4$ Levels and Implications for the Origin on Life

Much current origin of life research revolves around the "RNA world" hypothesis (Gilbert 1986; Joyce & Szostak 2018; Sasselov et al. 2020). This hypothesis proposes an interval of time when primitive life consisted of self-replicating, evolving RNA molecules, which, at some point, were encapsulated in cells. On a rocky world, "RNA world" requires that RNA is synthesized from early raw materials. Laboratory experiments that have successfully synthesized nucleobases, which are building blocks of RNA, require the following nitriles: hydrogen cyanide (HCN), cyanoacetylene (HCCCN), and cyanogen (NCCN; Sutherland 2016; Ritson et al. 2018; Benner et al. 2019). In addition, nitriles have also been used to synthesize amino acids (Miller & Urey 1959; Sutherland 2016).

The known natural source of nitriles is photochemistry in a chemically reducing atmosphere containing $H_2$, $CH_4$ and $N_2$ or perhaps $NH_3$. For example, Titan's photochemistry produces all the aforementioned nitriles (Strobel et al. 2009). Importantly, to make the simplest nitrile, HCN, requires abundant $CH_4$ because HCN is formed from photochemical products of $CH_4$ and nitrogen (Zahnle 1986; Tian et al. 2011).

Our results show that volcanic gases generally are unlikely to cause high atmospheric $CH_4$ abundances in prebiotic atmospheres. Consequently, the results lend credence to alternative proposals for creating early $CH_4$-rich, reducing atmospheres, such as impacts (Zahnle et al. 2020). Impacts can create a reducing atmosphere when reactions between iron-rich impact ejecta and shock-heated water vapor from an ocean generate copious $H_2$, $CH_4$, and $NH_3$. Subsequent photochemistry would generate HCN and other prebiotic nitriles over thousands to millions of years (Zahnle et al. 2020).

## 5. Conclusions

Our modeling of volcanic outgassing speciation suggests that chemically reducing volcanism on terrestrial planets is unlikely to mimic biological $CH_4$ fluxes. The improbable cases where volcanoes do produce biological $CH_4$ fluxes also often produce CO. Volcanoes are not prone to produce $CH_4$ for several reasons. First, the high solubility of $H_2O$ in magma limits the amount of total hydrogen outgassed, thus preventing the production of H-bearing molecules like $CH_4$. Second, $CH_4$ outgassing requires relatively low magma temperatures compared to the majority of magma erupted on Earth. Finally, $CH_4$ outgassing requires a very low magma oxygen fugacity, unlike that of most terrestrial planets inferred from astronomical data (Doyle et al. 2019).

We use a photochemical model to calculate the atmospheric composition of planets with volcanoes that produce $CH_4$. We find that atmospheric $CH_4$ should coincide with abundant CO. On the other hand, biogenic $CH_4$ can coincide with a low CO abundance if CO-consuming microbial life is present.

Therefore, the $CH_4$–$CO_2$ biosignature is most compelling when little or no atmospheric CO is detected. Atmospheric $CH_4$-$CO_2$ and large CO is ambiguous and can be explained by an uninhabited planet with highly reducing volcanic gases, or an inhabited planet where biology is unable to sequester atmospheric CO (Schwieterman et al. 2019).

However, observations of CO are not required to make conclusions about the abiotic or biotic origin of observed atmospheric $CH_4$. Atmospheric $CH_4$ and $CO_2$ alone would have a reasonable probability of being biological if the observed $CH_4$ abundance implies a surface flux similar to modern Earth's biological $CH_4$ flux (30 Tmol yr$^{-1}$). Such a large $CH_4$ flux is difficult to explain with reducing volcanic gases or other abiotic processes that generate $CH_4$, such as serpentinization.

These conclusions should be taken with caution because they are based on what is understood about processes occurring on the Earth and our Solar System, which may be a very sparse sampling of what is possible.

We thank Lena Noack, Michael McIntire, and the two anonymous reviewers for very knowledgeable and constructive comments and conversations. We also thank Max Galloway for pointing out a mistake in our calculations in an early draft of this article. N.W. and D.C.C. were supported by the Simons

**Table 3**
Mount Etna Magma Composition

| Magma component | Mole fraction |
|---|---|
| $x_{SiO_2}$ | 0.516 |
| $x_{TiO_2}$ | 0.014 |
| $x_{Al_2O_3}$ | 0.110 |
| $x_{FeO}$ | 0.091 |
| $x_{MgO}$ | 0.092 |
| $x_{CaO}$ | 0.126 |
| $x_{Na_2O}$ | 0.035 |
| $x_{K_2O}$ | 0.002 |
| $x_{P_2O_5}$ | 0.016 |

**Note.** Taken from Iacono-Marziano et al. (2012).

**Table 4**
Solubility Constants

| Constant | Value |
|---|---|
| $C_{CO_2}$ | 0.14 |
| $B_{CO_2}$ | −5.3 |
| $b_{CO_2}$ | 15.8 |
| $B_{H_2O}$ | −2.95 |
| $b_{H_2O}$ | 1.24 |
| $d_{Al_2O_3/(CaO+K_2O+Na_2O)}$ | 3.8 |
| $d_{FeO+MgO}$ | −16.3 |
| $d_{Na_2O+K_2O}$ | 20.1 |

**Note.** "Anhydrous" case from Iacono-Marziano et al. (2012).

Collaboration on the Origin of Life grant 511570 (to D.C.C.), as well as the NASA Astrobiology Program grant No. 80NSSC18K0829, and benefited from participation in the NASA Nexus for Exoplanet Systems Science research coordination network. J.K.-T. was supported by the NASA Sagan Fellowship and through the NASA Hubble Fellowship grant HF2-51437 awarded by the Space Telescope Science Institute, which is operated by the Association of Universities for Research in Astronomy, Inc., for NASA, under contract NAS5-26555.

## Appendix A
### Details of Outgassing Speciation Model

#### A.1. Solubility Constants for H$_2$O and CO$_2$

Our outgassing model uses solubility equations for H$_2$O and CO$_2$ in mafic magmas from Iacono-Marziano et al. (2012; Equations (1) and (2)). The parameters $S_1$ and $S_2$ in the solubility equations depend on the chemical make-up of the magma. We found that different mafic magma compositions did not significantly affect the outputs of our outgassing speciation model (Section 2.1); therefore, for the purposes of calculating melt solubility, we fixed the chemical make-up of the magma to the magma erupting at Mount Etna, Italy, reported by Iacono-Marziano et al. (2012). This reduced the complexity of the model without sacrificing any significant amount of accuracy.

Table 3 shows the chemical make-up of the magma at Mount Etna, and Table 4 shows several solubility constants from Iacono-Marziano et al. (2012). Together, these values define the solubility parameters $S_1$ and $S_1$:

$$S_1 = \ln\left(\frac{\mu_{magma}}{\mu_{CO_2} 10^6}\right) + \frac{C_{CO_2} P}{T} + B_{CO_2} + b_{CO_2}\left[\frac{NBO}{O}\right] \\ + \left(\frac{x_{Al_2O_3}}{x_{CaO} + x_{K_2O} + x_{Na_2O}}\right) d_{Al_2O_3/(CaO+K_2O+Na_2O)} \\ + (x_{FeO} + x_{MgO}) d_{FeO+MgO} + (x_{Na_2O} + x_{K_2O}) d_{Na_2O+K_2O}, \quad (A1)$$

$$S_2 = \ln\left(\frac{\mu_{magma}}{\mu_{H_2O} 10^2}\right) + \frac{C_{H_2O} P}{T} + B_{H_2O} + b_{H_2O}\left[\frac{NBO}{O}\right], \quad (A2)$$

$$\left[\frac{NBO}{O}\right] = \frac{2(x_{K_2O} + x_{Na_2O} + x_{CaO} + x_{MgO} + x_{FeO} - x_{Al_2O_3})}{2x_{SiO_2} + 2x_{TiO_2} + 3x_{Al_2O_3} + x_{MgO} + x_{FeO} + x_{CaO} + x_{Na_2O} + x_{K_2O}}. \quad (A3)$$

Here $T$ is magma temperature, $P$ is the total pressure of degassing, and $\left[\frac{NBO}{O}\right]$ is the amount of nonbridging oxygen per oxygen in the melt.

#### A.2. Derivation of Equations (11) and (12)

The following is a derivation for the atom conservation equation for carbon used in our outgassing model (Equation (11)). The derivation for the atom conservation equation for hydrogen follows the exact same procedure, so we do not include it.

Consider some volume of magma with gas bubbles in it that contains a total number of moles $\gamma_{tot}$. The total moles is the sum of the moles of magma ($\gamma_{magma}$), and the moles of gas in bubbles suspended in that magma ($\gamma_{gas}$):

$$\gamma_{tot} = \gamma_{gas} + \gamma_{magma}. \quad (A4)$$

Within this same volume of magma, the total moles of carbon ($\gamma_C^{tot}$) is equal to the moles of carbon in the gas phase ($\gamma_C^{gas}$) and the moles of carbon dissolved in the magma ($\gamma_C^{magma}$) combined:

$$\gamma_C^{tot} = \gamma_C^{gas} + \gamma_C^{magma}. \quad (A5)$$

We assume that the only carbon-bearing molecule that can dissolve in the magma is CO$_2$; therefore, $\gamma_C^{magma} = \gamma_{CO_2}^{magma}$. Dividing by $\gamma_{tot}$ and expanding gives

$$\frac{\gamma_C^{tot}}{\gamma_{tot}} = \frac{\gamma_{gas}}{\gamma_{tot}}\frac{\gamma_C^{gas}}{\gamma_{gas}} + \frac{\gamma_{magma}}{\gamma_{tot}}\frac{\gamma_{CO_2}^{magma}}{\gamma_{magma}}. \quad (A6)$$

We can replace $\frac{\gamma_{\text{magma}}}{\gamma_{\text{tot}}}$ with $1 - \frac{\gamma_{\text{gas}}}{\gamma_{\text{tot}}}$ using Equation (A4). This leaves us with

$$\frac{\gamma_C^{\text{tot}}}{\gamma_{\text{tot}}} = \frac{\gamma_{\text{gas}}}{\gamma_{\text{tot}}} \frac{\gamma_C^{\text{gas}}}{\gamma_{\text{gas}}} + \left(1 - \frac{\gamma_{\text{gas}}}{\gamma_{\text{tot}}}\right) \frac{\gamma_{CO_2}^{\text{magma}}}{\gamma_{\text{magma}}}. \quad (A7)$$

Here $\frac{\gamma_{CO_2}^{\text{magma}}}{\gamma_{\text{magma}}}$ is just $x_{CO_2}$ (the mol fraction of $CO_2$ in the magma; see Table 1). Also, we assume that $CO_2$, $CO$, and $CH_4$ are the only carbon-bearing gas species, so $\gamma_C^{\text{gas}} = \gamma_{CO_2}^{\text{gas}} + \gamma_{CO}^{\text{gas}} + \gamma_{CH_4}^{\text{gas}}$. Making substitutions gives

$$\frac{\gamma_C^{\text{tot}}}{\gamma_{\text{tot}}} = \frac{\gamma_{\text{gas}}}{\gamma_{\text{tot}}} \frac{\gamma_{CO_2}^{\text{gas}} + \gamma_{CO}^{\text{gas}} + \gamma_{CH_4}^{\text{gas}}}{\gamma_{\text{gas}}} + \left(1 - \frac{\gamma_{\text{gas}}}{\gamma_{\text{tot}}}\right) x_{CO_2}. \quad (A8)$$

Assuming the ideal gas law, $\gamma_i^{\text{gas}}/\gamma_{\text{gas}} = P_i/P$. Also, to make the equation more manageable, we substitute $\alpha_{\text{gas}} = \frac{\gamma_{\text{gas}}}{\gamma_{\text{tot}}}$, which is the total mols in the gas phase divided by the moles in the gas phase and magma combined:

$$\frac{\gamma_C^{\text{tot}}}{\gamma_{\text{tot}}} = \frac{P_{CO_2} + P_{CO} + P_{CH_4}}{P} \alpha_{\text{gas}} + (1 - \alpha_{\text{gas}}) x_{CO_2}. \quad (A9)$$

Magma sometimes freezes deep in the Earth as a glass before it releases any volatiles. Measurements of volatiles like $CO_2$ in such glasses are reported in terms of mass fractions (Wallace et al. 2015). To stay consistent with these unit conventions, we indicate the total carbon in un-degassed magma as a mass fraction of $CO_2$ ($m_{CO_2}^{\text{tot}}$). We can convert the mass fraction to a mole fraction using Equation (3):

$$\frac{m_{CO2}^{\text{tot}} \mu_{\text{magma}}}{\mu_{CO_2}} = x_{CO2}^{\text{tot}} = \frac{\gamma_{CO_2}^{\text{tot}}}{\gamma_{\text{tot}}} = \frac{\gamma_C^{\text{tot}}}{\gamma_{\text{tot}}}. \quad (A10)$$

Substituting Equation (A10) into Equation (A9) gives

$$\frac{m_{CO_2}^{\text{tot}} \mu_{\text{magma}}}{\mu_{CO_2}} = \frac{P_{CO_2} + P_{CO} + P_{CH_4}}{P} \alpha_{\text{gas}} + (1 - \alpha_{\text{gas}}) x_{CO_2}. \quad (A11)$$

Equation (A11) is identical to Equation (11).

### A.3. Graphite Saturation and the Solubility of CO, CH4, and H2

Several studies have shown that degassing can be affected by graphite saturation of magma (Hirschmann & Withers 2008) or by the solubility of CO, $CH_4$, and $H_2$ in magma (Hirschmann et al. 2012; Ardia et al. 2013; Wetzel et al. 2013). Our model for outgassing speciation used throughout the main text does not account for these complications. Here we show that our assumption is valid because it does not significantly change our results.

Consider the following equilibrium:

$$C + O_2 \leftrightarrow CO_2, \quad (A12)$$

$$K_9 = \frac{f_{CO_2}}{a_C f_{O_2}} \approx \frac{P_{CO_2}}{a_C f_{O_2}}. \quad (A13)$$

Here $K_9$ is the equilibrium constant given by $\exp(47457/T + 0.136)$, and $a_C$ is the activity of carbon. To incorporate graphite saturation into our model, we first calculate outgassing speciation using the model described in the main text (Section 2.1). Next, we check for graphite saturate by calculating the activity of carbon using Equation (A13). If $a_C < 1$, then we assume the melt is not graphite saturated and that the calculation is valid. If $a_C > 1$, then we assume graphite is saturated and recalculate outgassing speciation by replacing the carbon conservation equation (Equation (11)), with the graphite saturation equation with $a_C = 1$ (Equations (A13)). Here we are considering graphite saturation in the magma just before degassing occurs. Our treatment is different from, for example, the methods of Ortenzi et al. (2020) because they are accounting for graphite saturation much deeper in a planet during partial melting of the mantle.

Figure 10 is identical to Figure 3, except Figure 10 accounts for graphite saturation. Graphite saturation appears to have a small effect on the results; therefore, it is justified to ignore it.

To incorporate the solubility of $H_2$, $CH_4$, and $CO$ into our model, we add the following solubility relationships to or system of original outgassing equations (Section 2.1):

$$\exp(-11.403 - 0.000076P) = K_5 = \frac{x_{H_2}}{f_{H_2}} \approx \frac{x_{H_2}}{P_{H_2}}, \quad (A14)$$

$$\exp(-7.63 - 0.000193P) = K_6 = \frac{x_{CH_4}}{f_{CH_4}} \approx \frac{x_{CH_4}}{P_{CH_4}}, \quad (A15)$$

$$\exp(-41.02 - 0.00056P) = K_7 = \frac{x_{Fe(CO)_5}}{a_{Fe} f_{CO}^5} \approx \frac{x_{Fe(CO)_5}}{a_{Fe} P_{CO}^5}. \quad (A16)$$

Here pressure-dependent equilibrium constants $K_5$, $K_6$, and $K_7$ are from Hirschmann et al. (2012), Ardia et al. (2013), and Wetzel et al. (2013), respectively. For Equation (A16), we take the activity of iron to be $a_{Fe} = 0.6$, based on the experiments in Wetzel et al. (2013). Also, we only include the Equation (A16) when $f_{O_2} <$ IW-0.55 (IW is the iron-wustite mineral buffer) because Wetzel et al. (2013) only observed CO dissolved in magma for these low oxygen fugacities.

We also alter the carbon and hydrogen atom conservation equations to accommodate for new molecules in the melt.

$$\frac{m_{CO_2}^{\text{tot}} \mu_{\text{magma}}}{\mu_{CO_2}} = \frac{P_{CO_2} + P_{CO} + P_{CH_4}}{P} \alpha_{\text{gas}} + (1 - \alpha_{\text{gas}})(x_{CO_2} + x_{CO} + x_{CH_4}), \quad (A17)$$

$$\frac{m_{H_2O}^{\text{tot}} \mu_{\text{magma}}}{\mu_{H_2O}} = \frac{P_{H_2O} + P_{H_2} + 2P_{CH_4}}{P} \alpha_{\text{gas}} + (1 - \alpha_{\text{gas}})(x_{H_2O} + x_{H_2} + 2x_{CH_4}). \quad (A18)$$

Here $x_i$ is the mol fraction of species $i$ in the melt.

Figure 11 is identical to Figure 3, except Figure 11 accounts for $H_2$, $CH_4$, and CO solubility in magma. The solubility of these three molecules has a small effect on the results; therefore they can be ignored.

### A.4. Closed System Cooling and Chemical Kinetics

Our model for volcanic outgassing is a thermodynamic equilibrium model. We assume that during magma eruptions, gas bubbles chemically and thermally equilibrate with magma, and then they are released to the atmosphere unaltered (Figure 1). This does not exactly reflect real degassing.

In reality, the chemical composition of gas bubbles changes as bubbles leave the magma and enter the atmosphere

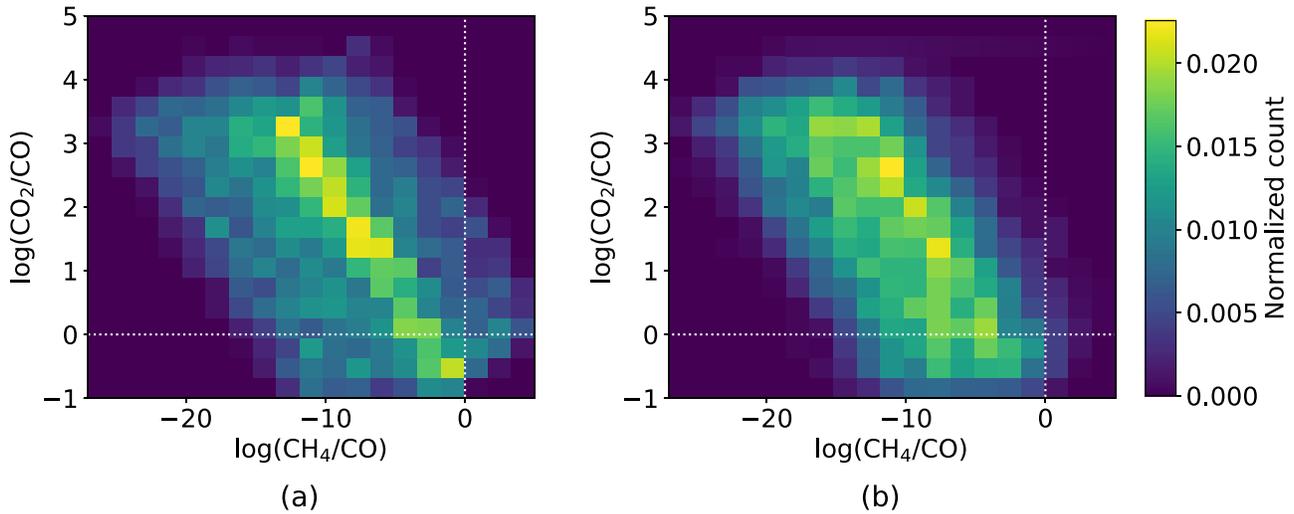

**Figure 10.** Identical to Figure 3, except here we account for graphite saturation in the melt. As in Figure 3, (a) is for ocean worlds and (b) is for Earth-like worlds. Graphite saturation has a small effect on the results.

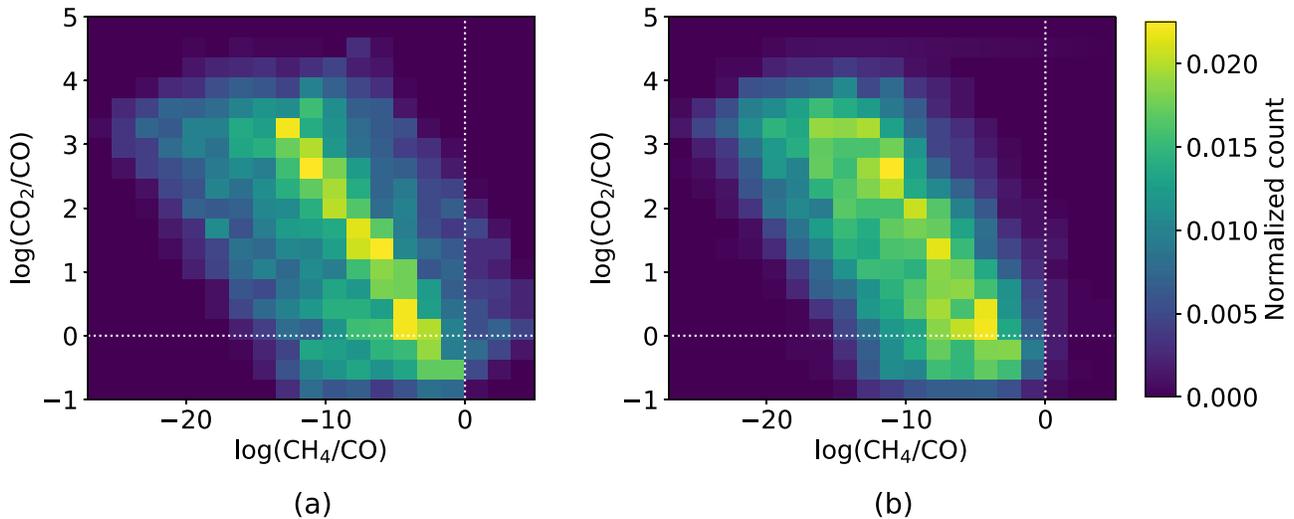

**Figure 11.** Identical to Figure 3, except here we account for the solubility of $H_2$, $CH_4$, and CO in the melt. As in Figure 3, (a) is for ocean worlds and (b) is for Earth-like worlds. $H_2$, $CH_4$, and CO solubility have a small effect on the results.

(Moussallam et al. 2019; Kadoya et al. 2020). As a bubble leaves magma, it cools down and new chemical equilibria are preferred. When a gas bubble first begins cooling, it is still very hot, so chemical reactions keep the bubble near chemical equilibrium. Once the bubble is cold enough, chemical reactions slow, and ultimately cease, quenching or freezing the chemical composition of the gas bubble. Therefore, the cooling process alters the chemistry of the gas.

Gas re-equilibration to lower temperatures explains the observed chemistry of volcanic gases globally (Moussallam et al. 2019), and Oppenheimer et al. (2018) provides a specific example of this phenomenon at in the Kilauea volcano in Hawaii. During eruptions at Kilauea, gas bubbles in the magma would rise to the surface. As the bubbles rose in the magma, they adiabatically expanded, which cooled the gas below the temperature of the magma. Chemical reactions during adiabatic expansion changed the chemical make-up of the bubble.

For the purposes of understanding potential $CH_4$ biosignature false positives from volcanoes, we need to know if bubble cooling might generate a substantial amount of $CH_4$. Here we first consider the kinetics of methane generation and show that reactions are likely too slow to generate substantial $CH_4$ during gas cooling. Next, we show that our Monte Carlo simulation results (Figure 4) remain qualitatively unchanged, even if our kinetics calculations are wrong, and $CH_4$ can be generated during gas cooling.

CO or $CO_2$ is converted to $CH_4$ through either of the net reactions (Schaefer & Fegley 2010):

$$CO + 3H_2 \leftrightarrow CH_4 + H_2O, \qquad (A19)$$

$$CO_2 + 4H_2 \leftrightarrow CH_4 + 2H_2O. \qquad (A20)$$

The rate-limiting step to either CO or $CO_2$ conversion to $CH_4$ is debated in the literature (Zahnle & Marley 2014), but the following are two solid candidates and their corresponding rate constants:

$$H_2 + H_2CO \rightarrow CH_3 + OH, \qquad (A21)$$

$$k_{10} = 2.3 \times 10^{-10} \exp(-36200/T), \qquad (A22)$$

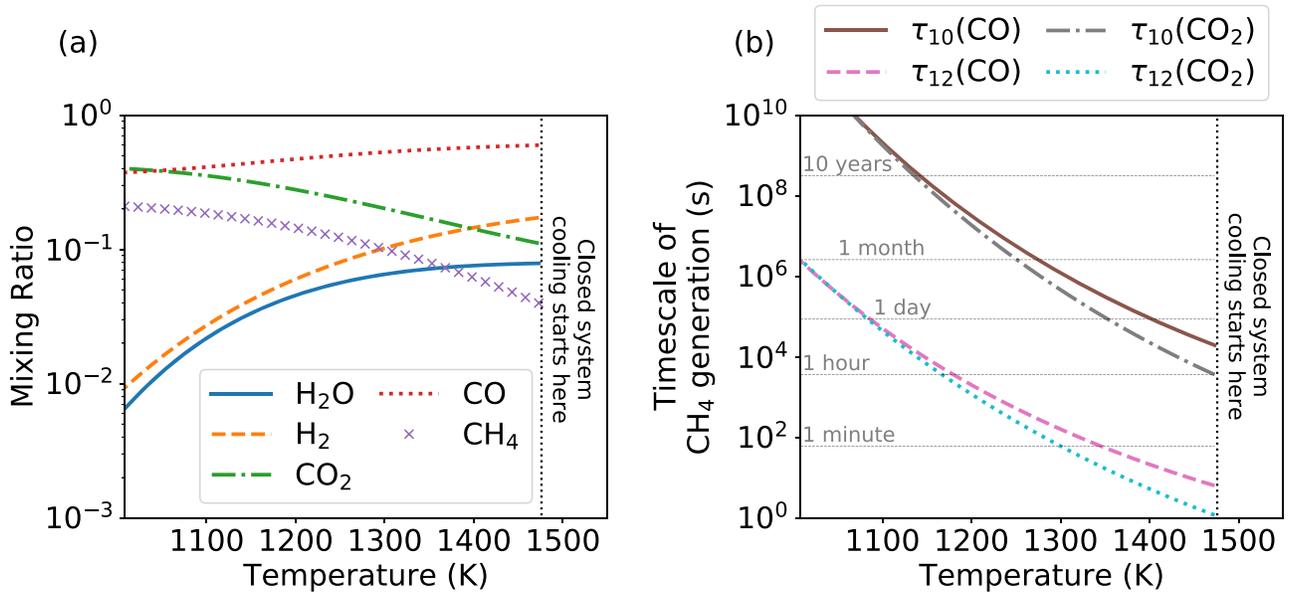

**Figure 12.** (a) Equilibrium composition as a function of temperature for a submarine volcanic gas that is cooled as a closed system and (b) timescales of CH$_4$ formation during closed system cooling. Timescales of volcanic gas cooling are not shown or calculated.

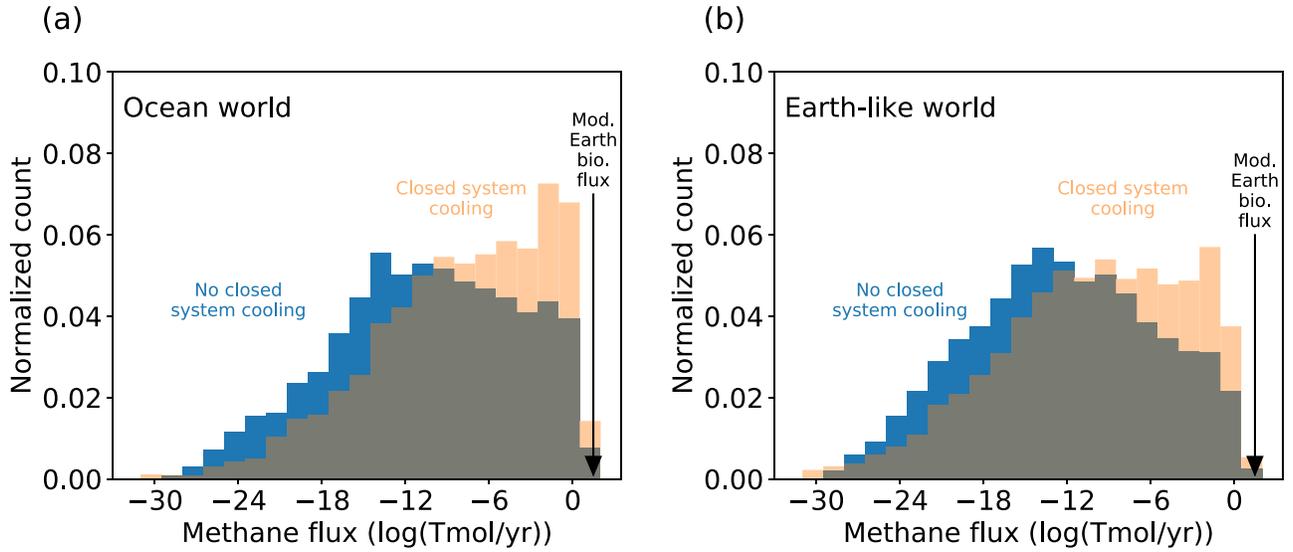

**Figure 13.** The blue histograms in (a) and (b) are identical to Figures 4(c) and (d), and orange histograms are identical Monte Carlo simulations, except they account for the closed system cooling of volcanic gases to equilibrium temperatures observed on Earth (800 to 1500 K). To calculate CH$_4$ fluxes, we used modern Earth's magma production rate.

$$\mathrm{H + H_2CO \rightarrow CH_3}, \qquad (A23)$$

$$k_{12} = 4.0 \times 10^{-11} \exp(-2068/T). \qquad (A24)$$

Here $k_{10}$ and $k_{12}$ are rate constants (cm$^3$ s$^{-1}$). The lifetime of CO or CO$_2$ conversion to CH$_4$ is thus one of the following:

$$\tau_{10}(\mathrm{CO}) = \frac{N_{\mathrm{CO}}}{k_{10} N_{\mathrm{H_2}} N_{\mathrm{H_2CO}}}, \qquad (A25)$$

$$\tau_{12}(\mathrm{CO}) = \frac{N_{\mathrm{CO}}}{k_{12} N_{\mathrm{H}} N_{\mathrm{H_2CO}}}, \qquad (A26)$$

$$\tau_{10}(\mathrm{CO_2}) = \frac{N_{\mathrm{CO_2}}}{k_{10} N_{\mathrm{H_2}} N_{\mathrm{H_2CO}}}, \qquad (A27)$$

$$\tau_{12}(\mathrm{CO_2}) = \frac{N_{\mathrm{CO_2}}}{k_{12} N_{\mathrm{H}} N_{\mathrm{H_2CO}}}. \qquad (A28)$$

Here $\tau$ is the chemical lifetime in seconds, and $N_i$ is the number density of species $i$ in molecules cm$^{-3}$.

Figure 12 shows timescales of CH$_4$ generation (Equations (A25)–(A28)) during the closed system cooling of submarine volcanic gas. To determine gas chemistry just before a bubble is released from magma, we use our speciation model (Section 2.1). At 1473 K, we calculate gas speciation assuming $P = 400$ bar, $f_{\mathrm{O_2}} = $ FMQ-4, $m_{\mathrm{CO_2}}^{\mathrm{tot}} = 0.1\%$, and $m_{\mathrm{H_2O}}^{\mathrm{tot}} = 0.5\%$. We then calculate new chemical equilibrium as the gas cools, assuming it is a closed system (i.e., we assume the gas is thermally and chemically decoupled from the magma; Figure 12(a)). Figure 12(b) shows the corresponding timescale of $CH_4$ generation (Equations (A25)–(A28)) at each temperature.

The quench temperature (i.e., the temperature where outgassing chemistry is frozen-in due to slow kinetics) of $CH_4$ depends on the cooling timescale of volcanic gases (not shown in Figure 12). $CH_4$ should quench where the cooling timescale is about the same as the timescale of $CH_4$ generation. After gases are released from a submarine volcano, we suspect they cool from magma temperatures to ocean temperatures on the order of seconds. If this is the case, then the $CH_4$ quench temperature is probably >1400 K. This would result in a negligible increase in the $CH_4$ content of the gas (Figure 12(a)).

Suppose that the $CH_4$ quench temperature was instead 1000 K. In this case, the $CH_4$ content of the gas would be increased by about a factor of five (Figure 12(a)). There are two ways that a ~1000 K $CH_4$ quench is possible. First, gas cooling could occur on timescales of months rather than seconds. According to Figure 12(b), month-long gas cooling should quench $CH_4$ by 1000 K. Second, catalysts could dramatically speed up the reactions creating $CH_4$, which might allow for quench temperatures near 1000 K for even gas cooling timescales of seconds. In the following two paragraphs, we show that either of these scenarios would not significantly change our results.

To demonstrate that re-equilibration of gases to feasible lower temperatures does not change our conclusions, assuming low $CH_4$ quench temperatures can be achieved, we perform another Monte Carlo simulation identical to the one described in Section 2.2, except we account for closed system cooling of volcanic gases. In the Monte Carlo simulation, we first calculate gas composition using our outgassing model (Section 2.1); then we re-equilibrate this gas mixture to the uniformly sampled gas equilibrium temperature between 800 and 1500 K. This range of gas equilibrium temperatures is the range observed in Earth's volcanic gases (Moussallam et al. 2019). In cases where the randomly drawn gas equilibrium temperature is higher than the magma temperature, we assume no closed system cooling occurs.

Figure 13 is identical to Figures 4(c) and (d), except Figure 13 accounts for closed system cooling of gases. Closed system cooling allows more $CH_4$ production on average, but still only 0.3% and 0.1% of calculations for ocean worlds or earth-like worlds, respectively, produce more than 10 Tmol $CH_4$ yr$^{-1}$. The probability of volcanic $CH_4$ fluxes being comparable to modern Earth's biological flux (30 Tmol yr$^{-1}$) is still low.

In summary, changes in gas chemistry during cooling might cause our speciation model to under-predict the $CH_4$ produced by an amount that does not change our conclusions significantly. Further consideration of the kinetics of $CH_4$ generation in volcanic gases is beyond the scope of this paper.

## Appendix B
## Photochemical Model Boundary Conditions

Table 5 shows boundary conditions used for the *Atmos* photochemical model. We used the same $H_2O$ and temperature profile as Kharecha et al. (2005) for all simulations. The version of Atmos that we used has updated rate constants and $H_2O$ cross sections following Ranjan et al. (2020).

Every simulation for planets orbiting the Sun uses a solar spectrum at 2.7 Ga, calculated via the methods described in

**Table 5**
Boundary Conditions for Photochemical Modeling

| Chemical species | Deposition velocity (cm s$^{-1}$) | Mixing ratio | Flux (molecules cm$^{-2}$ s$^{-1}$) |
|---|---|---|---|
| O | 1 | … | … |
| $O_2$ | $1.4 \times 10^{-4}$ | … | … |
| $H_2O$ | 0 | … | … |
| H | 1 | … | … |
| OH | 1 | … | … |
| $HO_2$ | 1 | … | … |
| $H_2O_2$ | $2 \times 10^{-1}$ | … | … |
| $H_2$ | 0 | … | $5.9 F_{CH_4}$ |
| CO | $1 \times 10^{-8}$ | … | $1.7 F_{CH_4}$ |
| HCO | 1 | … | … |
| $H_2CO$ | $2 \times 10^{-1}$ | … | … |
| $CH_4$ | 0 | … | variable |
| $CH_3$ | 1 | … | … |
| $C_2H_6$ | 0 | … | … |
| NO | $3 \times 10^{-4}$ | … | … |
| $NO_2$ | $3 \times 10^{-3}$ | … | … |
| HNO | 1 | … | … |
| $O_3$ | $7 \times 10^{-2}$ | … | … |
| $HNO_3$ | $2 \times 10^{-1}$ | … | … |
| $H_2S$ | $2 \times 10^{-2}$ | … | $0.1 F_{CH_4}$ |
| $SO_3$ | 0 | … | … |
| $S_2$ | 0 | … | … |
| HSO | 1 | … | … |
| $H_2SO_4$ | 1 | … | … |
| $SO_2$ | 1 | … | $0.1 F_{CH_4}$ |
| SO | 0 | … | … |
| $SO_4$ aerosol | $1 \times 10^{-2}$ | … | … |
| $S_8$ aerosol | $1 \times 10^{-2}$ | … | … |
| Hydrocarbon aerosol | $1 \times 10^{-2}$ | … | … |
| $CO_2$ | … | Variable | … |
| $N_2$ | … | 0.8 | … |

**Note.** Species included in the photochemical scheme with a deposition velocity and flux of 0 include N, $C_3H_2$, $C_3H_3$, $CH_3C_2H$, $CH_2CCH_2$, $C_3H_5$, $C_2H_5CHO$, $C_3H_6$, $C_3H_7$, $C_3H_8$, $C_2H_4OH$, $C_2H_2OH$, $C_2H_5$, $C_2H_4$, CH, $CH_3O_2$, $CH_3O$, $CH_2CO$, $CH_3CO$, $CH_3CHO$, $C_2H_2$, $(CH_2)_3$, $C_2H$, $C_2$, $C_2H_3$, HCS, $CS_2$, CS, OCS, S, and HS. Here deposition velocities follow those used by Schwieterman et al. (2019).

Claire et al. (2012), although our results are not sensitive to the age of the Sun. For planets orbiting an M8V star, we use estimates of TRAPPIST-1's spectrum derived by Lincowski et al. (2018), scaled so that the solar constant of the planet is 0.822 relative to modern Earth's. We use this solar constant because it places the simulated planet at the same relative distance from the inner edge of the habitable zone as Earth today (Kopparapu et al. 2013).

All of our models include the modern production rate of NO from lightning.

## ORCID iDs

Nicholas Wogan  https://orcid.org/0000-0002-0413-3308